\documentclass[11pt,a4paper]{article}
\usepackage{cite}
\usepackage{amsmath, amsthm, amssymb,slashed}
\usepackage{yfonts}
\usepackage[utf8]{inputenc}
\usepackage[english]{babel}
\usepackage{microtype}

\usepackage{ifpdf}
\ifpdf
 \usepackage[pdftex]{graphicx}
 \usepackage{epstopdf}
\else
 \usepackage[dvips]{graphicx}
\fi

\usepackage{subcaption}
\usepackage{xfrac} 
\usepackage{fancyhdr}
\usepackage{mathbbol}
\usepackage{url}
\usepackage{color}
\usepackage{bbm}
\usepackage{rsfso}
\usepackage{dsfont}
\usepackage{bm}
\usepackage{amsfonts}
\usepackage{mathrsfs}
\usepackage{slashed}
\usepackage{wasysym}
\usepackage{yfonts}
\usepackage{ytableau}
\usepackage{tikz}
\usepackage[hidelinks]{hyperref}
\usetikzlibrary{arrows,decorations.pathmorphing,hobby}
\usetikzlibrary{calc,arrows.meta}
\usepackage{cancel}
\usepackage[normalem]{ulem}%provides strikeout 

\def\bl{\begin{equation}\begin{aligned}}
\def\el{\end{aligned}\end{equation}}
\def\beal{\begin{align}}
\def\eal{\end{align}}
\def\be{\begin{equation}}
\def\ee{\end{equation}}
\def\bpm{\begin{pmatrix}}
\def\epm{\end{pmatrix}}
\def\bsm{\begin{bmatrix}}
\def\esm{\end{bmatrix}}
\def\bvm{\begin{vmatrix}}
\def\evm{\end{vmatrix}}
\def\bVM{\begin{Vmatrix}}
\def\eVM{\end{Vmatrix}}
\def\bea{\begin{eqnarray}}
\def\eea{\end{eqnarray}}

\def\1{{\bf 1}}
\def\2{{\bf 2}}
\def\3{{\bf 3}}
\def\4{{\bf 4}}

\def\hat{\widehat}

\usepackage{indentfirst}

\newcommand{\OMIT}[1]{}

\numberwithin{equation}{section}

\usepackage{mathtools}

\usepackage{fontawesome}
\usepackage{microtype,shapepar,xcolor}
\title{A short review on the compositeness of the $X(3872)$}
\date{\today}
\author{A. Esposito, A. Glioti, D. Germani and A.D. Polosa
\\ {\it $^*$Sapienza University of Rome and INFN, Piazzale Aldo Moro 2, I-00185, Italy} }
\begin{document}
\maketitle

\begin{abstract}
The $X(3872)$ could be a shallow $D\bar D^*$ bound state, a compact four-quark state, or a partially composite particle, i.e. a superposition of the two. We will review how these hypotheses could be tested experimentally, examining especially the cases in which the $X$ is a pure bound state or a pure compact tetraquark. Data on $X\to D\bar D\pi$ decays are compared with the analysis of the $X$ lineshape. The pure bound state hypothesis corresponds to a well-defined region in parameter space defined by the width of the $D^*$ versus the binding energy of the $X$. As for the $X$ lineshape, we observe that the currently available experimental analysis tests the compatibility with the compact hypothesis for the $X$. We propose how to extend the analysis to examine the molecular or the partially composite hypotheses. We also review the analysis on the radiative decays of the $X$ including pion corrections confirming some conclusions reached in the literature on the use of the universal wave function description for the molecular $X$. 
\end{abstract}

\newpage
\tableofcontents
\newpage

\section{Introduction}
The $D^0$ and $\bar D^{*0}$ mesons interact strongly at low energy, but the exact form of their interaction potential is unknown. If it were, we would know if a bound state is formed, together with its binding energy, $B$. The $S$-wave bound state would have $J^{P}=1^+$ quantum numbers.

In 2003 a very narrow $1^+$ resonance has indeed been discovered, the $X(3872)$, at a mass value which is almost exactly $m_D+m_{D^*}$~\cite{Belle:2003nnu}. It decays mostly in $D^0\bar D^0\pi^0$~\cite{ParticleDataGroup:2024cfk}, so that its identikit is seemingly simple: a bound state of $D$ and $\bar D^*$ with binding energy $B\simeq 0$ \cite{Braaten:2004fk,Lee:2009hy,Guo:2013zbw,Baru:2015nea}.\footnote{Throughout this work we often omit the neutral charge symbol, as well as the explicit expression for the charge conjugation eigenstate.} In low energy scattering theory this is known as a shallow bound state~\cite{DWBA}. Shallow bound states feature some universal properties, which are completely independent on the actual form of the binding potential. The deuteron is itself a shallow bound state of a neutron and a proton, $np$, with $B\simeq 2.2$~MeV, whereas in the case of the $X$ we have $B \simeq 100$~keV, or maybe less~\cite{Tomaradze:2015cza}. Thus the natural question: is the $X$ like the deuteron?

Despite the similarities highlighted above, the deuteron has never been observed in high-$p_T$ proton-proton collisions (but barely searched for), whereas the $X$ has a remarkable prompt production cross section for transverse momenta as large as $p_T\gtrsim 15$~GeV~\cite{CMS_pp_ptproduction,Esposito:2015fsa,LHCb_High_mult_pp_collision,CMS_pbpb_ptproduction}. It is difficult to imagine that such a loosely bound state is resilient enough to be produced as a metastable particle in prompt hadron-hadron collisions at high energy: when the typical momenta involved in the production process are much larger than the molecular binding momentum, it is extremely unlikely to form a shallow bound state with sizable cross section~\cite{prod1,prod2,prod3,prod4}. Along the same lines, a similar study of prompt production, as well as non-prompt production from beauty hadron decays, but in high-multiplicity $pp$ collisions, has shown a trend that does not match with a molecular structure~\cite{LHCb_High_mult_pp_collision,prompt_high_mult,non_prompt}.

A solution to the production problem may be found if it is assumed that the quantum state associated with the $X(3872)$, $|X\rangle$, which occurs in the production or decay amplitudes, is a superposition of an elementary $|\mathfrak X\rangle$ state, which here means a compact tetraquark, and a bound state $|\mathfrak{B}\rangle$ made of two open charm mesons $D\bar D^*$. This is found to help explain the large prompt production cross sections observed: the $X$ gets produced through its compact tetraquark component, which can also be copiously produced at high momenta.\footnote{Some studies distinguish between a compact $c\bar c $ core and a bound state~\cite{Meng:2013gga}. Here we will only consider the case in which $|\mathfrak X\rangle$ and $|\mathfrak B\rangle$ have the same quark content. }

We will discuss how the coupling of the bound state component, $\mathfrak B$, to its $D\bar D^*$ constituents, in a {\it partially composite} $X$, can be derived in the shallow bound state approximation. The coupling of the elementary $\mathfrak X$ to the $D\bar D^*$ continuum has to be treated, instead, as an additional parameter.

Consider the ideal case of performing a low-energy scattering experiment of $D$ and $\bar D^*$ particles. The pion exchange (in the $u$ channel) does not generate a binding potential \cite{Esposito:2023mxw}. If a $D\bar D^*$ bound state exists at all, it is generated by the short distance $(D\bar D^*)^2$ quartic interaction, which corresponds to a $\delta^3(\bm r)$ potential in the non-relativistic quantum theory. This potential, with a properly renormalized coupling, allows bound states~\cite{jackiw}. 

In non-relativistic quantum field theory, the formation of a molecular bound state manifests itself through the presence of a negative energy pole in the $D\bar D^*$ scattering amplitude. The latter, in turn, results from the summation of the bubble diagrams induced by the $(D\bar D^*)^2$ quartic interaction in the $s$-channel~\cite{Braaten:2015tga,Braaten:2020nmc}. In place of summing the bubble diagrams, we will use the complete propagator of some bare field $\Phi$ in the $s$-channel assuming a non-vanishing amplitude $\langle 0|\Phi(0)| \bm p_1,\bm p_2\rangle$ on the 2-particle (or higher) continuum and assuming that interactions could be strong enough to bind the 2-particle state. The spectral density in the K\"all\'en-Lehman (KL) representation of the complete propagator contains contributions from (free) multi-particle states, as well as from a single particle state if the $\langle 0|\Phi(0)| \bm p\rangle\equiv \sqrt{Z}$ amplitude is not zero. 

To achieve $\sqrt{Z}\neq 0$ one must introduce the field associated to the elementary $\mathfrak X$ in the Lagrangian. If a shallow bound state $|\mathfrak{B}\rangle$ in place of the continuum $|\bm p_1,\bm p_2\rangle=|D\bar D^*\rangle$ is introduced in the spectral representation, as done in Section~\ref{sbsqft}, a formula for the coupling of the bound state $\mathfrak B$ to $D\bar D^*$ is found, even in the presence of an elementary $\mathfrak X$ field in the Lagrangian. This formula is equivalent to what obtained by Weinberg~\cite{Weinberg:1965zz}\footnote{The same result can be obtained in an EFT framework, as done in \cite{WeinbergEFT}.} using other methods, reviewed in Section~\ref{zinqm}, and it includes as well a less known result by Landau~\cite{landau}, obtained exclusively in the case $Z=0$~\cite{Polosa:2015tra}.

If it were possible to establish from experiments whether $Z\neq 0$, with due accuracy, we would know if the $X$ has to be described as a pure molecule of open charm mesons, as a partially composite state or as a compact tetraquark. Extracting $Z$ from data seems an experimentally impossible task if $Z$ is understood as the normalization constant of a quantum field. However, following what was done by Weinberg for the deuteron~\cite{Weinberg:1965zz}, $\sqrt{Z}$ can alternatively be viewed as the amplitude of the compact $|\mathfrak X\rangle$ state within the physical state $|X\rangle$, see Section~\ref{zinqm}. This analysis can be pushed further by deriving the scattering length, $a$, and the effective range, $r_0$, as functions of $Z$. Therefore we can relate two experimentally measurable quantities, $a$ and $r_0$, to $Z$ (see Section~\ref{ar}). Under certain assumptions on the (undefinite) binding potential, the sign of $r_0$ alone is indicative of a molecule \cite{Weinberg:1965zz,Smorod,Bethe}, see Section~\ref{r0sign}, but the case of the $X(3872)$ still requires special care, as discussed in Sections~\ref{corrs} and~\ref{correpion}. 

The option of a purely elementary state, i.e. a compact tetraquark, is discussed partly in Section~\ref{laonly} and more extensively in Section~\ref{purelycompact}. In Section~\ref{laonly} we will assume the $X$ partial compositeness to study the $X\to D\bar D\pi$ decay, requiring the elementary component to have a negligible coupling to the continuum with respect to the the bound state --- whose coupling to $D\bar D^*$ is derived in the shallow bound state hypothesis. We will explore the parameter space under these conditions focusing especially on the boundary $Z=0$, corresponding to the purely composite hypothesis. The forbidden regions will be interpreted as corresponding to a sizeable coupling of the elementary component.

As illustrated in Section~\ref{flatt} and especially in Section~\ref{purelycompact}, the LHCb study \cite{LHCb:2020xds} is explicitly probing only the elementary nature of the $X$ but not its partial compositeness. This is because the lineshape fitting curve used in their analysis is obtained by an effective theory in which only the $\mathfrak XD\bar D^*$ coupling is different from zero, with no quartic coupling $(D\bar D^*)^2$, i.e. with no $\mathfrak B$ bound state included \cite{Artoisenet:2010va}. Therefore the goodness of the resulting fit is only an indication that the hypothesis of having just an elementary $\mathfrak X$ coupled to $D\bar D^*$ is compatible with data, but values of $Z$ derived by this parametrization are not meaningful to define a partial compositeness of the state. 

In order to study the molecular case, we propose to use an effective field theory in which the quartic coupling is included but no elementary field for the $X$ (see Sections~\ref{sec:pure_mol} and~\ref{sommariols}). In the same sections, it is reviewed also the role of coupled channels, which were not addressed in the original discussion by Weinberg.

The coupling of the $X$ to $D\bar D^*$ which can be extracted by the LHCb analysis, is the $\mathfrak X D\bar D^*$ coupling, and it should allow a good determination of the $X\to D\bar D\pi$ decay rate measured by other experiments, in the purely compact interpretation. We find that this self-consistency condition is indeed reached assuming a total width of the $D^*$ which is approximately 100~keV.

The claims on the evidence of a molecular $X$ are therefore not supported by observation.
The compact tetraquark interpretation appears to be less popular because the SU(3) methods that have been used to treat compact tetraquarks predict more states than observed~\cite{Maiani:2004vq}. In particular, they predict a second neutral $X^0$ and two charged partners $X^\pm $~\cite{Maiani:2017kyi}. Despite some attempts to find some selection rules~\cite{Esposito:2016itg,Esposito:2016noz}, this situation has essentially remained unchanged since the birth of the compact interpretation of the $X(3872)$ in 2004. The first claims about the molecular $(Z=0)$ interpretation insisted on the fact that the $X$ had to correspond to the $D^0\bar D^{*0}$ bound state only, with no charged partners. Only very recently, a study has shown that there should be charged partners in the molecular scenario as well, but in the form of virtual states~\cite{Zhang:2024fxy}. These charged partners have not been found, for the moment, not even in the $J/\psi\rho^\pm$ channel, where they were supposed to hide according to the compact tetraquark interpretation~\cite{Maiani:2017kyi}. It is also worth stressing that several other charged states have been found in the meanwhile, like the $Z(4430)$, which is remarkably far from threshold~\cite{LHCb:2014zfx}.
This has a clear interpretation in the compact tetraquark model~\cite{Maiani:2008zz1,Maiani:2008zz2}. 

\vspace{1em}

In our view, there is strong evidence against the purely molecular interpretation that comes from radiative decays of the $X$ in $\psi$ and $\psi^\prime$ particles~\cite{Grinstein:2024rcu,norarad}. It is found that the branching ratio of $X$ into $\psi^\prime \gamma$ is larger than that in $\psi \gamma$~\cite{LHCb_radiative}, in obvious contradiction with phase space arguments. In Section~\ref{raddec}, we remind that this experimental result is incompatible with a shallow bound state molecule\footnote{In \cite{Colangelo:2025uhs,DeFazio:2008xq,cc3andradiative,ccradiative1}, the radiative decay within the $c\bar{c}$ hypothesis is also investigated.}, where the $D$ and $\bar D^*$ mesons form a very large ($>10 $~fm) bound state~\cite{molecradiative,molecradiative1,molecradiative2}. The situation could be improved only at the price of making the sizes of open charm mesons much larger than what they are supposed to be. We also show that short-range pion interactions cannot deform the bound state universal wave function of the $X$ to the level of changing our conclusions. 

\vspace{1em}

Quite a few arguments reported in the following are textbook knowledge on low energy scattering theory and are reproduced and commented on to have a self-consistent presentation and to keep the discussion elementary. The main reference is the book by Landau and Lifshitz on Non-Relativistic Quantum Mechanics~\cite{lqm3-2}. This is particularly the case for Section~\ref{SA} and in part for Section~\ref{ar} and~\ref{zfls}.

We also stress that our review is truly meant to focus on the aspects related to the interplay between the compact and composite hypotheses for the structure of the $X(3872)$. For a more general and complete review of multi-quark states, the reader can refer, for example, to~\cite{Esposito:2016noz,Lebed:2016hpi,Guo:2017jvc,Brambilla:2019esw}.

\section{Low energy scattering and shallow bound states}
\label{SA}
Consider the motion of a particle of mass $m$ in a potential $V(r)$, vanishing at infinity more rapidly than $\exp(-Cr)$, with $C$ some constant. The scattering amplitude as a function of energy, $f(E)$, has no singularities in $E$ other than simple poles in correspondence of discrete energy levels (bound states). In particular, for negative energies close to some {\it shallow} bound state, $-B$, the amplitude takes the form,
\be
f(E)=-\frac{A_0^2}{2m}\frac{1}{E+B} \,.
\label{zero}
\ee
This scattering amplitude is universal: it does not depend on the details of $V$ featuring the shallow bound state. We give an argument for~\eqref{zero} following Landau~\cite{lqm3-2}. The asymptotic form of the wavefunction for the particle $m$ in $V(r)$, for any complex $E$, is
\be
\chi(r) = \mathcal{A}(E)\,\exp\left(-r\sqrt{-2mE}\right)+\mathcal{B}(E)\,\exp\left(r\sqrt{-2mE}\right) \,.
\label{uno}
\ee
Going from negative to positive real values of $E$, along a circular path, $\gamma$, in the upper complex half-plane, the complex variable $z=\sqrt{-E}$ undergoes a change in its argument given by $\Delta_\gamma \arg z=(0-\pi)/2$, and reaches the upper edge of the square root cut for $E>0$;
therefore along $\gamma$
\be
\left.\sqrt{-E}\right|_{E<0} \mapsto \exp(-i\pi/2)\sqrt{E}=-i\left.\sqrt{E}\right|_{E>0} \,.
\label{due}
\ee
The real part, $\Re \sqrt{-E}$, is positive everywhere on the physical sheet, which corresponds to the complex $z$ plane excluding a cut along the positive real axis. 
Inserting~\eqref{due} in~\eqref{uno} gives, for $E>0$,
\be
\chi(r)=\mathcal{A}(E)\,\exp(ikr)+\mathcal{B}(E)\,\exp(-ikr) \,,
\label{tre}
\ee
with\footnote{\label{foot1} The result in~\eqref{tre} is different if the path $\gamma$ is taken in the lower half plane since, in that case, $\Delta_\gamma \arg z=(2\pi-\pi)/2=\pi/2 $. Therefore, in place of~\eqref{due}, we have that 
$\sqrt{-E}\big|_{E<0}\mapsto \exp(i\pi/2)\sqrt{E}=i\sqrt{E}\big|_{E>0}$, so that $\chi(r)=\mathcal{A}(E^*)\,\exp(-ikr)+\mathcal{B}(E^*)\,\exp(ikr)$. The reality of $\chi$ requires from~\eqref{uno} that $\mathcal{A}(E^*)=\mathcal{A}^*(E)$ and $\mathcal{B}(E^*)=\mathcal{B}^*(E)$, and the fact that $\chi$ has to be single-valued requires $\mathcal{A}^*(E)=\mathcal{B}(E)$.} 
\be
k=\sqrt{2mE} \,.
\ee
The coefficients $\mathcal{A}(E)$ and $\mathcal{B}(E)$ are non-singular everywhere except for the branch point at $E=0$. 
The presence of a {\it bound state} at $E=-E_0~(E_0>0)$ implies for the (reduced) wavefunction $\chi$ in~\eqref{uno} to vanish at infinity. Therefore it is required that 
\be
\mathcal{B}(E)=\mathcal{B}(-E_0)+\beta\big(E-(-E_0)\big)+\dots=\beta(-E_0)(E+E_0)+\dots \,,
\label{Bexpand}
\ee
with $\beta$ a constant, so that 
\be 
\label{noing} \mathcal{B}(-E_0)=0 \,.
\ee 
Consider now a {\it low energy} scattering, where $E$ is positive but small. If the binding energy, $E_0$, is also small (i.e. a {\it shallow} bound state), then we can still approximate,
\be
\mathcal{B}(E)\simeq \beta(E+E_0) \,.
\label{sette}
\ee
Comparing~\eqref{tre} with the phase-shifted reduced wave function 
\be
\chi(r)=C\sin(kr+\delta_0)=2iC\left(e^{i(kr+\delta_0)}-e^{-i(kr+\delta_0)}\right) \,,
\label{otto}
\ee
we immediately deduce the phase of the wavefunction to be,\footnote{In this paper we are considering solely the $S$-wave case, $\ell=0$. If $\ell\neq 0$ the wavefunction~\eqref{otto} is instead~\cite{lqm3-2}, $$\chi(r)=C\left(e^{i(kr-\ell\frac{\pi}{2}+\delta_\ell)}-e^{-i(kr+\ell\frac{\pi}{2} +\delta_\ell)}\right)\,,$$ and the term $\mathcal{A}/\mathcal{B}$ gets substituted by $(-1)^\ell \mathcal{A}/\mathcal{B}$.\label{foot2}}
\be
\frac{\mathcal{A}(E)}{\mathcal{B}(E)}=-\exp(2i\delta_0) \,.
\label{nove}
\ee
Using the general scattering amplitude formula  and formulae~\eqref{due} and~\eqref{nove} we get
\begin{align}
f&=\frac{1}{2ik}\big(\exp(2i\delta_0)-1\big)=\frac{-1}{2i\sqrt{2 m E}}\left(\frac{\mathcal{A}}{\mathcal{B}}+1\right)=\frac{1}{2\sqrt{-2 m E}}\left(\frac{\mathcal{A}}{\mathcal{B}}+1\right) \,.
\label{otto1}
\end{align}
Analytically continuing the scattering amplitude to values in the proximity of $E\simeq -E_0$, using~\eqref{sette}, we have 
\begin{align}
f(E)&\simeq \frac{1}{2\sqrt{2 m E_0}}\frac{\mathcal{A}(-E_0)}{\beta(E+E_0)}\equiv\frac{A_0}{2\beta\sqrt{2 m E_0}}\frac{1}{(E+E_0)} \,.
\end{align}
We will show soon that
\be
\beta=-\frac{1}{A_0}\sqrt{\frac{m}{2E_0}} \,,
\label{beta}
\ee
which proves~\eqref{zero}. Before discussing this latter result, observe that $A_0$ also defines the normalization of the reduced bound state wavefunction 
\be
\chi(r) =A_0\exp\left(-r\sqrt{2mE_0}\right) \,,
\label{universalreduced}
\ee
which can be found immediately to be
\be
A_0=(8mE_0)^{1/4} \,.
\ee
This gives the formula that we will mostly use in the rest of this work,
\be
f(E)=-\sqrt{\frac{2B}{m}}\frac{1}{E+B}
\label{principale} \,,
\ee
where we defined \be B\equiv E_0 \,.\ee
Now for the derivation of $\beta$ in~\eqref{beta}. Consider the reduced Schr\"odinger equation
\be
\chi^{\prime\prime}+2m(E-V)\chi=0 \,,
\ee
and its derivative with respect to energy
\be
\Big(\frac{\partial}{\partial E}\chi\Big)^{\prime\prime}+2m \chi+ 2m(E-V)\frac{\partial}{\partial E}\chi=0 \,.
\ee
Multiply the first by $\partial\chi/\partial E$ and the second by $\chi$ and subtract the second from the first to obtain, upon integration over $r$,
\be
\chi^\prime\frac{\partial \chi}{\partial E}-\chi
\left(\frac{\partial \chi}{\partial E}\right)^\prime=2m\int_0^r dr\, \chi^2(r)\to 2m\quad \text{as}\quad r\to\infty \,.
\label{der}
\ee
In the left-hand side of~\eqref{der} use the function $\chi$ as given by the asymptotic form 
\be
\chi(r)=\mathcal{A}(E)\,\exp\left(-r\sqrt{-2mE}\right)+\mathcal{B}(E)\,\exp\left(+r\sqrt{-2mE}\right) \,,
\ee
as in~\eqref{uno}. In the vicinity of $E=-B$ it reads
\be
\chi(r)=A_0\,\exp\left(-r\sqrt{2mB}\right)+(E+B)\,\exp\left(+r\sqrt{-2mB}\right) \,.
\label{chix}
\ee
Inserting~\eqref{chix} into~\eqref{der} we have
\be
-2A_0\beta\sqrt{2mB}=2m \,.
\ee
which is equivalent to~\eqref{beta}. 

\section{Shallow bound states in quantum field theory}
\label{sbsqft}
The scattering amplitude formula derived in Section~\ref{SA},
\be
f(E)=-\sqrt{\frac{2B}{m}}\frac{1}{E+B} \,,
\label{main}
\ee
is valid for {\it low energy scattering} in the presence of a {\it shallow bound state}.
This can be related to the scattering of two particles having reduced mass $m$ in the initial state $\Psi_{\alpha}$ into the same two particles, in the final state $\Psi_\beta$. We will refer to this scattering amplitude as $f(\alpha\to \beta)$. 

In the formalism of non-relativistic quantum mechanics particles are neither created nor destroyed. Therefore if the state $\Psi_\alpha$ describes a $D$ and a $\bar D^*$ mesons, the same ones must also be present in the state $\Psi_\beta$. The state corresponding to a free $D\bar D^*$ pair, with some relative momentum $\bm p$ in the center of mass, has an amplitude
\be
(\Psi_\alpha, V \Psi)=(\Psi_\beta, V \Psi)=g_{B} 
\label{g0}
\ee 
with the state $\Psi$ of the bound $D\bar D^*$ pair, with momentum $p=p_D+p_{\bar D^*}$ and mass $m=m_D+m_{D^*}-B$, assuming that some binding interaction is at work. If the bound state is very shallow, as in the case of the $X$, the state $\Psi$ is almost indistinguishable from the continuum states $\Psi_\alpha$ or $\Psi_\beta$. 

Let us find an expression for $f(\alpha\to \beta)$ at low energy introducing the propagation of a composite, bound state as in the following quantum field theory derivation,\footnote{\label{footnote5}The square root in the first line of formula~\eqref{f_alpha_beta} contains the phase space factor $\delta^4(p_\alpha-p_\beta)d\beta\to k^\prime E_1^\prime E_2^\prime/E \, d\Omega $ and the flux factor $1/u_\alpha$ where $u_\alpha=kE/E_1E_2$. The definition of $M_{\beta\alpha}$ in Weinberg-I,~Eq.~(3.4.10)~\cite{Weinberg_QFT} has a $(2\pi)^3$ inside with respect to the one used here. The overall phase in the definition of $f$ is conventional. Observe that $f$ must have the dimensions of a length, therefore we must have that $[g_{B}^2]=1/E$, as it turns out to be the case in~\eqref{landau}.}
\begin{align}
f(\alpha\to \beta) &=\frac{1}{2\pi }\sqrt{
\frac{k^\prime E_1^\prime E_2^\prime E_1E_2}{k E^2}} \, M_{\beta\alpha}=
\frac{1}{8\pi E}(2m_D)(2m_{D^*})\, M_{\beta\alpha}\notag \\
& =\frac{1}{8\pi E}(2m_D)(2m_{D^*})(2m_X)
\, M_{\beta X}\, \Delta^\prime(p) \, M_{X\alpha}\,,
\label{f_alpha_beta}
\end{align}
where $\Delta^\prime (p)$ is the complete propagator of some field $\Phi$, which can be elementary or composite, in the K\"all\'en-Lehman (KL) form~\cite{Peskin_Sch},
\be 
\Delta^\prime(p)=\int\frac{\sigma(\mu^2)}{p^2+\mu^2-i\epsilon}d\mu^2 \,,
\label{klsig}
\ee
as obtained by setting $Z=0$ in the spectral function $\rho(\mu^2)=Z\, \delta(\mu^2-m^2)+\sigma(\mu^2)$. 
The following Lehman sum rule holds
\be
1=Z+\int\sigma(\mu^2)\, d\mu^2 \,,
\ee
where $Z$ is the renormalization constant of the bare field $\Phi$ whose propagartor is given in~\eqref{klsig} and $0\leq Z\leq1$.\footnote{If $|\bm p\rangle$ is a one-particle state with mass $m$ with a non-zero amplitude
$\langle 0|\Phi(0)|\bm p\rangle\neq 0$, Lorentz invariance requires
\be
\langle 0|\Phi(0)|\boldsymbol p\rangle=\frac{N}{\sqrt{2E}} \,, \quad \text{ with } \quad E=\sqrt{\boldsymbol p^2+m^2} \,. \notag
\label{ops}
\ee
Then, according to a general result \cite{Peskin_Sch}, the complete propagator $\Delta^\prime(p)$ of the bare field $\Phi$ has a pole at $-m^2$ with residue 
\be
Z=|N|^2>0 \,,
\ee
so that 
\be
\Delta^\prime (p)=\frac{Z}{p^2+m^2-i\epsilon} \,,
\label{58}
\ee
consistently with the KL formula. Now, in a generic, relativistic quantum field theory $Z$ has no probabilistic interpretation, contrary to the description in Section~\ref{zinqm}. However, a strictly non-relativistic quantum field theory is in 1-to-1 correspondence with standard non-relativistic quantum mechanics, meaning that they predict the same observables. In this instance, the probabilistic interpretation of $Z$ is recovered.} If $Z>0$ the bare field $\Phi$ appears in the Lagrangian and therefore the particle associated with it has to be considered as {\it elementary}. Here, instead, we assume $Z=0$ because we consider low energy non-relativistic $D\bar D^*$ scattering in a theory with only $D$ and $D^*$ mesons and without an elementary $X$. 
The spectral function $\rho(\mu^2)$ is defined by,\footnote{The complete propagator $\Delta^\prime(p)$ is the Fourier transform of $\langle 0 |T\Phi(x)\Phi^\dag(y)|0\rangle$ where $\langle0|\Phi(x)\Phi^\dag(y)|0\rangle$ is expressed as a sum over a complete set of states, $\langle0|\Phi(x)\Phi^\dag(y)|0\rangle=\sum_n \langle0|\Phi(x)|n\rangle\langle n|\Phi^\dag(y)|0\rangle$, and similarly for $\langle0|\Phi^\dag(y)\Phi(x)|0\rangle$. 
\label{comple2}}
\be
\theta(p_0)\, \rho(-p^2)=\sum_n\delta^4(p-p_n)|\langle 0|\Phi(0)|n\rangle|^2 \,,
\ee 
and $|n\rangle$ represents single or multi-particle states as in $|n\rangle=|\bm p\rangle, |\bm p_1,\bm p_2\rangle, \dots$. The single particle state is responsible for the Dirac-delta term in the spectral density\footnote{Considering the contribution from the one-particle state only~\eqref{ops}
\be
\theta(p_0)\, \rho(-p^2)=\sum_n\delta^4(p-p_n)|\langle 0|\Phi(0)|n\rangle|^2=\int\delta^4(p-p_1)\frac{Z}{2E}d^3 p_1+\dots \,,
\ee
and
\be
\int\delta^4(p-p_1)\frac{Z}{2E}d^3 p_1=
Z\int d^4p_1\theta(p_{10})\delta(p_1^2+m^2)\delta^4(p-p_1)=Z\theta(p_{10})\delta(p_1^2+m^2) \,,
\ee
so that in our case
\be
\rho(\mu^2)=Z\delta(\mu^2-m^2_X) \,.
\ee
} while the multiparticle states make the $\sigma(\mu^2)$. 
If $Z=0$ we have 
\be
1=\int\sigma(\mu^2)\, d\mu^2 \,,
\ee
which can be solved in our case by 
\be
\sigma(\mu^2)=\delta\big(\mu^2-(m_D+m_{D^*})^2\big) \,,
\ee
assuming that the $|D\bar D^*\rangle$ state is made up of non-interacting particles, as in the KL formalism. On the other hand when we take $Z=0$ means that the in-state couples as strongly as possible to the continuum $D\bar D^*$: the elementary fields in the Lagrangian ($D,D^*$) have so strong mutual interactions to change the spectrum of states from $|n\rangle$ as described above. This rules out the use of perturbation theory in the quantum field theory description. We assume that a shallow bound state is formed non-perturbatively and truncates the contribution of higher multiparticle states. We assume 
\be
\sigma(\mu^2)=\delta\big(\mu^2-(m_D+m_{D^*}-B)^2\big) \,,
\label{introB}
\ee
with $B\simeq 0$. Therefore we get from~\eqref{f_alpha_beta}
\begin{align}
f(\alpha\to \beta) &=\frac{1}{2\pi }\sqrt{
\frac{k^\prime E_1^\prime E_2^\prime E_1E_2}{k E^2}} \, M_{\beta\alpha}=
\frac{1}{8\pi E}(2m_D)(2m_{D^*})\, M_{\beta\alpha}\notag \\
& =\frac{1}{8\pi E}(2m_D)(2m_{D^*})(2m_X)
\, M_{\beta X} \frac{1}{p^2+m_X^2-i\epsilon} M_{X\alpha} \notag\\&= \frac{1}{8\pi E} 8 m m_X^2 \frac{ (\Psi_\beta, V \Psi) (\Psi, V \Psi_\alpha) }{p^2+m_X^2-i\epsilon} \notag\\
&= \frac{1}{8\pi E} 8 m m_X^2 \frac{ g_{B}^2}{p^2+m_X^2-i\epsilon} \,,
\label{deri}
\end{align}
where 
\be
p_\mu=(p_D+p_{\bar D^*})_\mu \,,
\ee
and 
\be
m_X=m_D+m_{\bar D^*}-B \,.
\ee
Thus, we get,
\begin{align}
(p_D+p_{D^*})^2+m_X^2&=(p_D+p_{D^*})^2+(m_D+m_{D^*}-B)^2 \notag\\
&\simeq \underbrace{(\boldsymbol p_D+\boldsymbol p_{D^*})^2}_{\simeq 0}-(m_D+m_{D^*}+\underbrace{\frac{p_D^2}{2m_D}+\frac{p_{D^*}^2}{2m_{D^*}}}_{E})^2+(m_D+m_{D^*}-B)^2\notag \\
&\simeq- 2(m_D+m_{D^*})\,(E+B)\simeq -2m_X\,(E+B) \,,
\end{align}
neglecting terms of order $B^2$ (shallow bound state) and $E^2$ (scattering at low energy). Defining $$g^2=8 m m_X^2 g_{B}^2\,,$$ we obtain Landau's result in the same form as presented in~\cite{landau},
\be
f(\alpha\to \beta) =\frac{1}{8\pi m_X}\frac{ g^2}{(p_D+p_{D^*})^2+m_X^2-i\epsilon}\simeq 
-\frac{1}{16\pi m_X^2}\frac{ g^2}{E+B+i\frac{\epsilon}{2m_X} } \,.
\label{f1}
\ee
If the width of the $X$ is taken into account, the infinitesimal $\epsilon$ must be replaced by a finite $\gamma$. Recalling that $\gamma$ is related to the total width $\Gamma$ by $\Gamma=\gamma/m_X$ we get 
\be
f(\alpha\to \beta) =-\frac{m}{2\pi}\frac{ g_{B}^2}{E+B+i\frac{\Gamma}{2} } \,.
\label{f11}
\ee
Neglecting the total width (a good approximation in the case of the $X$~\cite{LHCb:2020xds}),
\be
f(\alpha\to \beta) =-\frac{m}{2\pi}\frac{ g_{B}^2}{E+B} \,.
\label{f111}
\ee
Formula~\eqref{f111} has to be compared with the low energy scattering amplitude in the presence of shallow bound states~\eqref{main}
\be
f(\alpha\to \beta)=-\sqrt{\frac{2B}{m}}\frac{1}{E+B} \,.
\label{f2}
\ee
In this description, the $X$ is not a new particle being propagated between two interaction vertices. In the spectrum of the non-relativistic theory we assume to have only $D$ and $\bar D^*$ and indeed we are treating $X$ as a $D\bar D^*$ bound state, with very small $B$. 
The comparison between the two formulas in~\eqref{f111} and~\eqref{f2} gives immediately
the coupling $g_{B}$ in~\eqref{g0} in the form 
\be
g^2_{B}=\frac{2\pi}{m} \, \sqrt{\frac{2B}{m}}
\label{landau}
\ee
This formula for the amplitude was first found by Landau~\cite{landau}, not making any reference to the KL formalism but just introducing the propagator as in~\eqref{deri}.
Weinberg obtained independently\footnote{In a way not related to the work done by Landau. } a similar formula in the form~\cite{Weinberg:1965zz}
\be
g^2_Z=\frac{2\pi}{m} \, \sqrt{\frac{2B}{m}} (1-Z) \,,
\label{weinberg}
\ee
which is the same as~\eqref{landau} if $Z=0$. The extra $Z$ term is related to the one-particle contribution in the KL propagator $\Delta^\prime(p)$ associated with the elementary state. The combination $(1-Z)$ in formula~\eqref{weinberg} is due to the use of the Lehman sum rule. 
Indeed taking into account a $Z\neq 0$, and solving the Lehman sum rule as done above,
\be
\Delta^\prime(p)=\int\frac{\sigma(\mu^2)}{p^2+\mu^2-i\epsilon}d\mu^2=\frac{1-Z}{p^2+(m_D+m_{D^*}-B)^2-i\epsilon}= \frac{1-Z}{p^2+m_X^2-i\epsilon} \,.
\label{klderi}
\ee
Considering that $g_{B}$ is the coupling to the two-particle state we get in place of~\eqref{deri} 
\be
f(\alpha\to \beta)=- \frac{m}{2\pi}\frac{g_{B}^2\, (1-Z)}{E+B}=-\sqrt{\frac{2B}{m}}(1-Z)\frac{1}{E+B} \,.
\label{516}
\ee
In the same way as the residue at the pole of~\eqref{f2} is defined to be $(m/2\pi)\, g_{B}^2$, the residue at the pole of~\eqref{516} is defined to be $(m/2\pi)\, g_Z^2$. This suggests to introduce the coupling of the bound state $\mathfrak B$ to $D\bar D^*$ given by~\eqref{weinberg} 
\be
g^2_Z=\frac{2\pi}{m} \, \sqrt{\frac{2B}{m}} (1-Z) \,,
\label{328}
\ee
or equivalently having
\be
\frac{A_0^2}{2m}=\sqrt{\frac{2B}{m}}(1-Z) \,,
\label{a0bas}
\ee
rather than
\be
\frac{A_0^2}{2m}=\sqrt{\frac{2B}{m}}\equiv \frac{1}{mR_0} \,,
\label{330}
\ee
where 
\be
R_0=\frac{1}{\sqrt{2mB}} \,.
\ee
The $g_Z^2$ here is the coupling of the scattering particles $D$ and $D^*$ to the bound state $\mathfrak B$, but, in the presence of an elementary state in the Lagrangian, which gets reduced by $(1-Z)$ with respect to the Landau $g_{B}^2$.
So we can say that $g_Z^2$ is the {\it coupling of the scattering particles to the $\mathfrak B=D\bar D^*$ bound state, in the presence of an elementary state}.

The coupling to the bound state is zero if $Z=1$: we are suppressing the multiparticle contribution in the propagator and so there is no $B$ introduced in~\eqref{deri} and consequently the formula for $g_Z^2$ in~\eqref{weinberg} is not even defined, given that it depends on $B$. The case of $Z=1$ could be summarized by saying that the $X$ is elementary with an effective coupling $g_c$ to $D\bar D^*$ (different from $g_{Z=0}=0$, and at this level unspecified).

In the next Section, we will derive the $(1-Z)$ factor following Weinberg, with a non-relativistic quantum mechanics approach. For $Z\neq0$ the spectrum has to be extended from $D,\, \bar D^*$ to $D,\, \bar D^*,\, \mathfrak{X}$. Here $\mathfrak{X}$ is not a bound state of $D$ and $\bar D^* $, but rather a particle with a different structure, as elementary as $D$ and $\bar D^*$ are. In the amplitude in~\eqref{g0}, $\Psi$ has to be replaced by a superposition of $\mathfrak{X}$ and the continuum $D\bar D^*$ (see~\eqref{quella} in the next Section), meaning that the spectrum has been enlarged to include $D,\, \bar D^*,\, \mathfrak{X}$, all elementary at the same level. If $Z=0$, the spectrum is reduced to $D,\, \bar D^*$ only. 

\section{$Z$ in quantum mechanics}
\label{zinqm}
In Weinberg's derivation of formula~\eqref{weinberg} it is assumed that the quantum state describing the physical $X$ particle is a superposition of the quantum state describing an elementary $\mathfrak{X}$ and the $D\bar D^*$ bound state $\mathfrak{B}$ as in~\cite{Weinberg:1965zz},
\be |X\rangle=\sqrt{Z}|\mathfrak{X}\rangle+|\mathfrak{B}\rangle \,,
 \label{quella}
\ee
where 
\be
|\mathfrak{B}\rangle= \int_{\boldsymbol p}\, C_{\boldsymbol p} |D\bar D^*(\boldsymbol p)\rangle \,,
\label{quattrodue}
\ee
and $\bm p$ is the relative momentum of the $D\bar D^*$ pair. The wavefunction of the bound state, $\langle \bm x|\mathfrak B\rangle=\Psi(\bm x)$, is determined by the $C_{\bm p}$ components, and corresponds to the universal shallow bound state wavefunction given, in the reduced form, in~\eqref{universalreduced}, or in the standard form in~\eqref{universal}.
The completeness relation is assumed
\be
1=|\mathfrak{X} \rangle \langle\mathfrak{X}|+\sum_{\alpha} |\alpha\rangle \langle \alpha|
\label{comple1}
\ee
where by $|\alpha\rangle$ we mean the continuum $|\alpha\rangle=|\bm p_1,\bm p_2\rangle$, and $\bm p_i$ are the momenta of the $D$ mesons. The orthogonality $\langle \alpha|\mathfrak X\rangle=0$ implies $\langle \mathfrak X|\mathfrak B\rangle=0$. 
We might name~\eqref{quella} as the partial elementarity of the $X$, or conversely its {\it partial compositeness}.
Requiring the normalization 
\be
\langle X|X\rangle =1
\ee
is equivalent to $\langle\mathfrak{B}|\mathfrak{B}\rangle=1-Z$ with $\langle \mathfrak X|\mathfrak X\rangle=1$. Namely $Z$ is the probability that an examination of the state describing $X$ will find it in the elementary particle state rather than in the two-particle {\it bound} state.

We will now show how to obtain the relation
 \be
g_Z^2= |\langle D \bar D^* |V|\mathfrak{B} \rangle |^2=\frac{2\pi}{m}
\sqrt{\frac{2 B}{m}}(1-Z) \,,
\label{wei}
 \ee
 which represents the coupling of the bound state $\mathfrak{B}$ to the scattering states in the presence of an elementary $\mathfrak{X}$, and due to an unspecified potential $V$. The interaction $V$ generates bound states at a negative energy $-B$ from the meson-meson threshold, i.e.,
\be
H |\mathfrak{B}\rangle=(H_0+V)|\mathfrak{B}\rangle=-B|\mathfrak{B}\rangle \,,
\label{acca0pv}
\ee 
where 
\be
H_0|\alpha\rangle=E(\alpha)|\alpha\rangle \,.
\label{acca0}
\ee
From $\langle X|X\rangle=1$ and the completeness relation~\eqref{comple1} we get
\be
1=\langle X| \mathfrak{X}\rangle\langle\mathfrak{X}| X\rangle +\sum_\alpha \langle X|\alpha\rangle\langle\alpha |X\rangle=Z+\int |\langle\alpha|\mathfrak{B}\rangle|^2 \, d\alpha \,,
\ee
since $\langle \alpha|\mathfrak{X}\rangle=0\label{zeroampele}$. 
Therefore we obtain 
\be
1-Z=\int |\langle \alpha|\mathfrak{B}\rangle|^2 d\alpha=\int \frac{ |\langle \alpha|V|\mathfrak{B}\rangle |^2}{(E(\alpha)+B)^2} d\alpha \,,
\label{umz}
\ee
where the last equality can be understood by writing
\be
V=(H_0+V)-H_0 \,,
\ee
and using Eq.~\eqref{acca0pv} and~\eqref{acca0}. 
In the case $Z=0$, Eqs.~\eqref{umz} corresponds to
\be
1=\int |\langle \alpha|\mathfrak{B}\rangle|^2\, d\alpha \,,
\ee
or $|\langle \mathfrak{B}|\mathfrak{B}\rangle|^2=1$, otherwise we use
\be
1-Z=\int \frac{ |\langle \alpha|V|\mathfrak{B}\rangle |^2}{(E(\alpha)+B)^2} d\alpha \,.
\label{umz1}
\ee
For very small $B$, this is how the shallow bound state hypothesis comes in: this integral nearly diverges and it can be approximately evaluated restricting to low-energy $|\alpha\rangle$ states, corresponding to small $E(\alpha)$ values, by replacing 
\be
|\langle \alpha|V|\mathfrak{B}\rangle |^2\longrightarrow g_Z^2~\text{(constant)} 
\ee
in the energy range $0\leq E\leq\Lambda$. As in the derivation based on the K\"all\'en-Lehman relation in~\eqref{klderi}, the $g_Z^2$ coupling represents the coupling of the bound state $\mathfrak B=D\bar D^*$ to the $D\bar D^*$ scattering state in the presence of an elementary state $\mathfrak X$. The $g_Z^2$ coupling gets reduced by $(1-Z)$ with respect to the $g_{B}^2$ coupling of the pure bound state to the continuum $D\bar D^*$. We also have to replace the integration measure by the non-relativistic two particle state energy integral (since $E(\alpha)\sim B\sim 0$) 
\be
d\alpha=\frac{d^3 p}{(2\pi)^3}=\frac{p^2 dp}{2\pi^2} =\frac{1}{2\pi^2} \big(\sqrt{2m E}\big)^2\frac{d\sqrt{2m E}}{dE} dE=\frac{1}{4\pi^2}(2m)^{3/2}\sqrt{E} dE \,.
\label{eq:phase_space}
\ee
Therefore we should compute 
\be
\int_0^\Lambda\frac{\sqrt{E}}{(E+B)^2}dE=\frac{\tan ^{-1}\left(\sqrt{\frac{\Lambda }{B}}\right)}{\sqrt{B}}-\frac{\sqrt{\Lambda
 }}{B+\Lambda } \,.
\label{415}
\ee
As for $\Lambda$, we are assuming $\Lambda > B$. In the original analysis by Weinberg, the size of the deuteron, $1/\sqrt{2mB}$, is larger than the range of pion-exchange interactions. Indeed, with $B=2.2$~MeV we get $1/\sqrt{2mB}\simeq 4.3$~fm whereas $1/m_\pi\simeq 1.46$~fm. The condition\footnote{According to what found in~\eqref{virial}, $1/\sqrt{2mB}$ corresponds to the (reduced) De Broglie wavelength of the molecule.} $1/\sqrt{2mB}>1/m_\pi$ is equivalent to $\Lambda>B$ if we simply assume 
$\Lambda=m_\pi^2/2m$ where $m$ is the reduced mass of the $np$ system, so that $\Lambda\simeq 20$~MeV, consistent with $\Lambda\gg B$. With this approximation in mind we can rather compute 
\be
\int_0^\infty\frac{\sqrt{E}}{(E+B)^2}dE=\frac{\pi}{2\sqrt{B}} \,.
\label{infi}
\ee
and readily derive~\eqref{wei}, from formula~\eqref{umz1}. Consistently it turns out that $g_Z$ is very small if $B\to 0$. 

As we will see in Section~\ref{corrs}, the case of the $X$ is more subtle, and $m_\pi$ must rather by replaced by a scale $\mu$, such that 
$1/\mu=4.6$~fm. Moreover, $B$ is anywhere between $0$ and $0.3$~MeV (see Section~\ref{laonly}). For this latter value $1/\sqrt{2mB}\simeq 8.17$~fm. Here $\Lambda=\mu^2/2m\simeq 0.8$~MeV which is $\Lambda\gg B$ for small values of $B$ only. Anyway we can go beyond the approximation in~\eqref{infi} using~\eqref{415} and finding 
\be
g_Z^2= \frac{\sqrt{2} \pi ^2 (1-Z)}{m^{3/2} \left(\frac{\tan ^{-1}\left(\sqrt{\frac{\Lambda
 }{B}}\right)}{\sqrt{B}}-\frac{\sqrt{\Lambda }}{B+\Lambda }\right)} \,,
\label{wei2}
\ee
which in the $\Lambda\gg B$ case is equivalent to~\eqref{wei}. 
 
The case $Z=0$ has a clear interpretation: the $|X\rangle$ particle is purely a $D\bar D^*$ bound state. The case $Z=1$, instead, implies the absence of the bound state, as also implied by the vanishing of $g_Z$. In this case, the $|X\rangle =|\mathfrak X\rangle$ state can still decay into $D\bar D^*$, but through the amplitude pertaining solely to the elementary states,
\be
\langle \alpha|V|\mathfrak X\rangle \equiv g_c\neq g_Z \,,
\label{418}
\ee
which was not included in the analysis done above. This coupling can be extracted from the Flatt\'e LHCb analysis reviewed in Section~\ref{zfls}. 

\vspace{1em}

The only conclusion that can be drawn from Weinberg's analysis of the deuteron~\cite{Weinberg:1965zz} is that it spends less than approximately 10$\%$ of its time in a compact deuteron state (using data available in~\cite{Weinberg:1965zz}). We will discuss in Section~\ref{raddec} that in the case of the $X(3872)$ it is instead the compact state to be prominent.
The same conclusion is obtained by studying the lineshape of the $X$, as in Section~\ref{purelycompact}. In both cases this is done without having to extract $Z$ directly from data. 

\vspace{1em}

Finally we observe that the coefficients $C_{\bm p}$ in~\eqref{quattrodue} might not correspond to the Fourier components of the bound state wavefunction of a $D\bar D^*$ pair. Consider, in fact, the instance of no bound state, when the $X$ is a purely compact state $\mathfrak X$. We define $| X^{(+)}\rangle$ as its scattering in-state, the one to be used when computing the scattering matrix elements. Considering that, by definition of in-state, 
\be
(H_0+V)|X^{(+)}\rangle=E_X |X^{(+)}\rangle \,,
\ee
and
\be
H_0|\mathfrak X\rangle=E_X|\mathfrak X \rangle \,,
\ee
we get~\cite{DWBA},
\be
|X^{(+)}\rangle=|\mathfrak X\rangle+\frac{1}{E_X-H_0+i\epsilon}V|X^{(+)}\rangle \,.
\ee
Using the completeness relation~\eqref{comple1}, and 
$\langle \mathfrak X|V|X^{(+)}\rangle=\langle \mathfrak X|E_X-H_0|X^{(+)}\rangle=0$, we get the following expression for the in-state, 
\be
|X^{(+)}\rangle=|\mathfrak X\rangle +\int_\alpha |\alpha\rangle \frac{\langle \alpha|V|X^{(+)}\rangle}{E_X-E_\alpha+i\epsilon} \,.
\ee
The $|\alpha\rangle$ state depends on two continuous variables: the two momenta of the $D$ and the $\bar D^*$, as in $|\alpha\rangle =|\bm p_1,\bm p_2\rangle$. 
Again we require the $|X^{(+)}\rangle$ to be normalized $\langle X^{(+)}|X^{(+)}\rangle=1$. In order to also normalize $\langle \mathfrak X|\mathfrak X\rangle=1$, we can introduce 
\be
|X^{(+)}\rangle=\sqrt{Z}|\mathfrak X\rangle +\int_\alpha |\alpha\rangle \frac{\langle \alpha|V|X^{(+)}\rangle}{E_X-E_\alpha+i\epsilon}
\ee
so that using $\langle \mathfrak X|\alpha\rangle =0$ one has 
\be
1=Z+\int_\alpha \frac{|\langle \alpha|V|X^{(+)}\rangle|^2}{(E_X-E_\alpha+i\epsilon)^2} \,.
\ee
This time $Z$ is related to the dressing of the $\mathfrak X$ particle due to interactions with $D\bar D^*$, which do not form any bound state. Therefore, it is not always correct to state that $Z$ measures the partial compositeness of the $\mathfrak X$ which, in this case, is solely elementary. The case discussed in Section~\ref{purelycompact} corresponds to this latter case as long as quartic interactions $(D\bar D^*)^2$ are neglected . On the other hand quartic interactions can give rise to a bound state as in~\eqref{quella} | where the dressing of the $\mathfrak X$ is still included in the definition of $Z$. The pure bound state case is discussed in Section~\ref{sec:pure_mol}.

\section{$Z$ from the decay $X\to D\bar D\pi$}
\label{laonly}
\label{gzexp}

In this section we wish to compute the coupling $X\to D\bar D\pi$, under the hypothesis that $0\leq Z<1$ and that (see~\eqref{418})
\be
g_c\ll g_Z \,.
\ee
In Section~\ref{purelycompact} we will study, instead, the opposite instance, in which $g_c$ is the prominent coupling.

For the moment we are assuming that there is a compact component in the $X$ (with possibly small $Z$) weakly coupled to the $D\bar D^*$ continuum so that the coupling of the physical $X$ to the continuum is given by,
\be
g_X^2 = {|\langle \alpha | V | X \rangle |}^2 = {|\sqrt{Z} g_c + g_Z|}^2 \simeq g_Z^2 \,,
\ee
or equivalently 
\be
g_X^2\simeq \frac{2\pi}{m} \, \sqrt{\frac{2B}{m}} (1-Z) \,.
\label{53}
\ee
If the above expression for $g^2_X$ were found not to describe data, for example if ones finds $Z<0$ from experiment, then the only conclusion than can be drawn is that a relevant part of the dynamics is described by the compact component~\cite{Polosa:2015tra}, and by the compact coupling $g_c$; yet we cannot obtain both $g_c$ and $Z$ from the single decay process we want to analyze. The partial decay rate is~\cite{Polosa:2015tra} 
\be
\frac{d\Gamma(X\to D\bar D\pi)}{ds}=2 g^2\, \frac{p(m_X^2,m_D^2,s)}{8\pi m_X^2}\, \frac{1}{\pi}\frac{(s/m_{D^*} )\, {\cal B}(D^*\to D\pi)\Gamma_{D^*}}{(s-m_{D^*}^2)^2+((s/m_{D^*})\Gamma_{D^*})^2}\,\frac{\frac{p(s,m_D^2,m_\pi^2)}{2\sqrt{s}}}{\frac{p(m_{D^*}^2,m_D^2,m_{\pi}^2)}{2m_{D^*}}} \,,
\label{deca}
\ee
where\footnote{The coupling $g^2$ has dimensions $[g^2]=E^2$ so it gives the width $\Gamma=p/(8\pi m_X^2)\times g^2$, $p$ being the decay momentum.} we are assuming
\be
g^2= 8mm_X^2 \, g_X^2\simeq 8mm_X^2 \, g_Z^2 \,,
\ee
as in~\eqref{deri}, but taking $g_Z$ in place of $g_{B}$. The coupling $g$ defines the amplitude
\be
\langle D^0\bar D^{*0}|X\rangle=g\, e_X\cdot e_{D^*}^* \,,
\ee
which we assume to be dominated by the bound state component. Here $e_X^\mu$ and $e_{D^*}^\mu$ are the 4-vector polarizations of the $X$ and the $D^*$. In addition we used the fact that 
\be
\frac{1}{3}\sum_{\rm pol.}|e_X\cdot e_{D^*}^*|^2\simeq 1 \,.
\ee
In~\eqref{deca} we defined 
\be
p(m_1^2,m_2^2,m_3^2)=\frac{\sqrt{\lambda(m_1^2,m_2^2,m_3^2)}}{2m_1} \,,
\ee
with
\be
\lambda(x,y,z)=x^2+y^2+z^2-2xy -2xz -2yz \,.
\ee
The factor of 2 in formula~\eqref{deca} comes from the $1/\sqrt{2}$ in the $C=+1$ final state $(|D\bar D^*\rangle+|\bar DD^*\rangle)/\sqrt{2}$. 

Using the current determination \cite{ParticleDataGroup:2024cfk} of the branching fraction, ${\cal B}(X\to D\bar D\pi)=0.45\pm0.21$, 
the central mass values $m_{D}=1864.84$~MeV, $m_{D^*}=2006.85$, $m_X=3871.65$, the widths
$\Gamma_X= 1.19\pm0.21$~MeV, $\Gamma_{D^{*0}}\simeq 2$~MeV (at the upper bound of the value reported in the PDG), and $B=m_D+m_{D^*}-m_X$, we get from the partial rate in~\eqref{deca} (with $s_{\rm min}=(m_D+m_\pi)^2$ and $s_{\rm max}=(m_X-m_D)^2$),
\be
Z=0.1\pm 0.4 \,,
\label{primaz}
\ee
where the error derives from the ones in ${\cal B}(X\to D\bar D\pi)$ and $\Gamma_X$. 
For all $Z<0$ unacceptable values we simply conclude that the formula for the coupling of $X$ to $D\bar D^*$, based on the partial compositeness hypothesis of a shallow bound state and of a very weak coupling of the elementary component, cannot be applied: a sizeable $g_c$ should be included. If this is done, there is an entire region of the $(g_c,Z)$ parameter space giving the $g_X$ needed to reproduce data. 

\vspace{1em}

This result shows clearly how important it is to diminish the uncertainty on the total width and on the branching ratio of the $X$. A precise determination of $B$ is also essential: we report here for example the determination $B=3\pm192$~KeV, as discussed in~\cite{Tomaradze:2015cza}: with a $B=195$~KeV we would have had $Z=0.71\pm0.15$, for example. 

\begin{figure}
 \centering
 \includegraphics[width=0.5\textwidth]{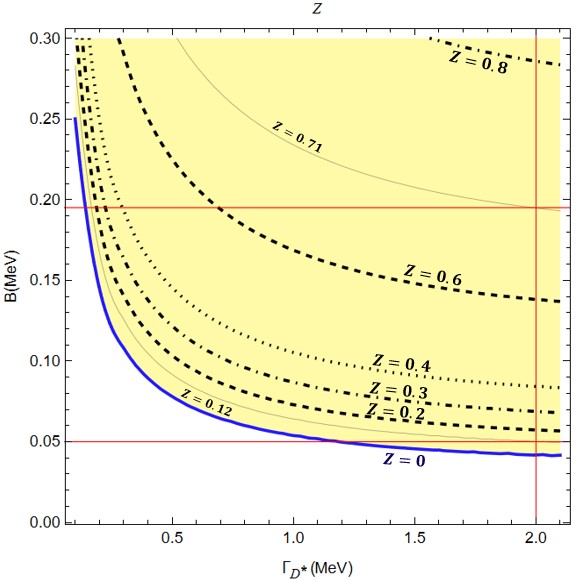}
    \caption{\small Using $g_X^2\simeq g_Z^2$ as in~\eqref{wei2}. The shadowed region represents $Z>0$ values in the $g_Z^2$ formula for the partially composite $X$. On the $x$ axis the total $D^*$ width. On the $y$ axis the binding energy $B$. Uncertainties reported on the PDG \cite{ParticleDataGroup:2024cfk} have been taken into account. The pure molecule case corresponds only to the $Z=0$ blue line. The unshadowed areas require sizeable values of the $g_c$ coupling.}
\label{fig1}
\end{figure}

The shadowed region in Fig.~\ref{fig1} is the one in which $Z\geq 0$ values are obtained. In the unshadowed region the small $g_c$ hypothesis fails, pointing to an important (or even exclusive) role of the compact component. Instead, the $Z=0$ case, which corresponds to the blue line in Fig.~1, does not depend on $g_c$. 

\vspace{1em}

Of course we should also consider the opposite instance, not considered here: the one in which $g_c$ is dominant, or it is the only coupling ($Z=1$), i.e. the case in which the $X$ corresponds to a purely elementary particle state. This case will be studied in Section~\ref{purelycompact}. 
It is found there that using LHCb data on the $X$ lineshape, the coupling $g_c$ can be determined under the hypothesis $Z=1$. 
Assuming that the $X$ is entirely elementary, this $g_c$ value can be used to determine the $X\to D\bar D\pi$ decay rate finding agreement with data for a perfectly reasonable value of $\Gamma_{D^{*0}}\simeq 100$~keV. 

We conclude that it is quite difficult, with the data at hand, to study the partially composite case, but the two extreme cases $Z=0$ and $Z=1$ can be more easily addressed. In particular it turns out the the purely elementary case is very well supported by the available data on the lineshape of the $X$, whereas the data on its radiative decays, see Section~\ref{raddec}, tend to exclude the $Z=0$ case. 

\section{$Z$ from the effective range and scattering length}\label{ar}
Let us go back to the scattering of two slow particles, $kR\ll1$, described by an attractive potential, $V$, with range $R$, featuring a shallow bound state at the discrete level $-B$. With a different reasoning with respect to the one illustrated in Section~\ref{SA}, it can be shown that the $S$-wave phase shift is determined by the relation
\be
\cot\delta=-\sqrt{\frac{B}{E}} \,.
\label{phaseshift}
\ee 
Let us see briefly why. In region II ($r>R$), outside the potential range, the particles are approximately non-interacting, with a reduced wave function $\chi_{\rm II}(r)\propto \sin(kr+\delta)$. Given that we assume $kR\ll1$, $\chi_{\rm II}(r)$ varies slowly as $r\to 0$. 
Because of the slow variation of $\chi_{\rm II}(r)$, the matching condition $(\chi^\prime/\chi)_{\rm II}=(\chi^\prime/\chi)_{\rm I}$, to be taken at some $r^*>0$, could equally be computed at $r^*=0$. Therefore we obtain $(\chi^\prime/\chi)_{\rm II}=k\cot\delta$. 

Within region I ($r<R$), the Schr\"odinger equation will not depend explicitly on energy, as $V\gg B$ for a shallow bound state, and the boundary condition will not depend on the total energy either. Given that, we are free to choose the energy as that of a stationary state, $E =-B$. In this case $\chi_{\rm I} \propto e^{-\kappa r}$, with $\kappa=\sqrt{2m B}$. Therefore, using the latter: $(\chi^\prime/\chi)_{\rm I}=-\kappa$. Since $k$ in region II is $k=\sqrt{2mE}$, the boundary condition at $r^*$ is $\cot\delta=-\sqrt{B/E}$. 

This implies that the general formula for the scattering amplitude can be written as $(k=\sqrt{2mE})$
\be
f(\alpha\to \beta)=\frac{1}{k\cot\delta-ik}=-\frac{1}{\sqrt{2m}}\frac{\sqrt{B}-i\sqrt{E}}{E+B} \,,
\label{otheruni}
\ee
which again does not depend on the explicit form of the potential $V$: it is {\it universal}. 
This same formula and the phase shift in~\eqref{phaseshift} can be alternatively derived with the aid of the Low equation, in the case of shallow bound states~\cite{Weinberg:1965zz,DWBA}.

In Section~\ref{SA} we have discussed how $\sqrt{-E}$ gets transformed into $-i\sqrt{E}$, and viceversa, in moving towards positive values of $E$ through the upper half of the complex plane ($\sqrt{-E}~\big|_{E<0} \rightleftarrows -i\sqrt{E}\big|_{E>0}$ or, alternatively, $\sqrt{2mB}\rightleftarrows -ik $). Therefore, in the vicinity of $E=-B$, the previous formula in~\eqref{otheruni} gets transformed into
\be
f(\alpha\to \beta)=-\frac{1}{\sqrt{2m}}\frac{2\sqrt{B}}{E+B} \,,
\ee
which corresponds to
\be
f(\alpha\to \beta)=-\sqrt{\frac{2B}{m}}\frac{1}{E+B} \,,
\ee
as found in~\eqref{principale}. A scattering length, $a$, can be introduced
\be
k\cot \delta=-k\sqrt{\frac{B}{E}}=-k\sqrt{\frac{B}{k^2/2m}}=-\sqrt{2mB}\equiv-\frac{1}{a} \,.
\label{eq:a_B}
\ee
Notice that this relation between $a$ and $B$ hold only at this order in the small $k$ expansion. When including further subleading terms, it will have to be updated.
From the equation above, one can again deduce the universal formula for low energy scattering, which matches the shallow bound state formula,
\be
f(\alpha\to \beta)=\frac{1}{-1/a-ik}= \frac{-1/a+ik}{1/a^2 + k^2} = -\sqrt{\frac{2B}{m}}\frac{1}{E+B} \,,
\ee
as $B=1/2ma^2$. To obtain this, we used the fact that, generically, $k^2 = 2mE$ and $1/a=\sqrt{2mB}$. Moroever, in the limit of small $E$ and $B$ that we are working on, we can replace $E\simeq-B$ in the numerator, which is regular, i.e. $ik \simeq - \sqrt{2mB}$.

A {\it refined } expansion at low energies --- keep the next term at low energy beyond $1/a$ --- gives\footnote{The generic amplitude $$f(E)=\frac{1}{g(E)-ik}$$ is a real function in the vicinity of $E=-B$ being $g(E)$ real and $ik$ real as well. Therefore it can be expanded in powers of $E$, i.e. in \emph{even} powers of $k$ --- this is why in~\eqref{r0exp} the linear term is missing. }
\be
k\cot\delta\simeq -\frac{1}{a}+\frac{1}{2}r_0 k^2 \,,
\label{r0exp}
\ee
where $r_0$ is the so called effective range. Therefore
\be
f(\alpha\to \beta)=\frac{1}{-1/a+\frac{1}{2}r_0k^2-ik} \,.
\label{err0bethe}
\ee

The pole corresponding to the shallow bound state, at $E=-B$, appears at $k=i\sqrt{2m B}\equiv i\varkappa$, as found solving~\cite{Esposito:2021vhu} 
\be
\left(-\varkappa_0 +\frac{1}{2}r_0k^2- ik\right)_{k=i\varkappa}=0 \,,
\label{cond}
\ee
where $\varkappa_0=1/a$. The latter condition~\eqref{cond} gives
\be
-\varkappa_0=-\varkappa+\frac{1}{2}r_0\varkappa^2 \,,
\ee
which can be plugged back into Eq.~\eqref{err0bethe}. Following similar manipulations as above one gets, for a generic $k=\sqrt{2mE}$,
\begin{align}
f={}&\frac{1}{\frac{r_0}{2}(k^2+\varkappa^2)-(\varkappa +ik)}=\frac{1}{\frac{r_0}{2}(k^2 +\varkappa^2) - \frac{k^2+\varkappa^2}{\varkappa - i k}} = \frac{1}{\frac{r_0}{2} - \frac{1}{\varkappa - i k}} \frac{1}{2m(E+B)} \notag \\
={}& \frac{1}{\frac{r_0}{2} - \frac{1}{2\varkappa}} \frac{1}{2m(E+B)} = -\frac{1}{m(R_0-r_0)}\frac{1}{E+B} \,,
\label{err0formw}
\end{align}
where $R_0=\frac{1}{\sqrt{2mB}}$, and again we replaced $k \simeq i \varkappa$ in the regular prefactor. This means that 
\be
\frac{A_0^2}{2m}=\frac{1}{m(R_0-r_0)} \,,
\label{a0r0}
\ee
in place of~\eqref{330},
\be
\frac{A_0^2}{2m}=\frac{1}{mR_0} \,.
\ee
So, expanding the scattering amplitude to include the effective range, $R_0$ must be shifted by $r_0$ in the formula for the residue at the shallow bound state pole $A_0^2/2m$. 
The latter gives Landau's coupling $g_{B}^2$. The residue at the pole $A_0/2m$, corresponds to $(m/2\pi) \, g_Z^2$ as given in~\eqref{a0bas} \footnote{It follows that $Z=0\Rightarrow r_0=0$. }
\be
\frac{A_0^2}{2m}=\frac{1}{mR_0}(1-Z)=\frac{1}{m(R_0-r_0)} \,,
\label{614}
\ee
finding 
\be
r_0=-\frac{Z}{1-Z}R_0\,.
\label{err0}
\ee
Using the expression for $r_0$ just obtained in the pole condition Eq.~\eqref{cond} we obtain the {\it positive} scattering length (remember that $Z\leq 1$)
\be
a_0=\frac{2(1-Z)}{2-Z}R_0 \,.
\label{aa}
\ee
The quantities $r_0$ and $a_0$ are both experimentally measurable from the low momentum limit of the scattering amplitude. Formulae~\eqref{err0} and~\eqref{aa} are the main results of what several authors call {\it Weinberg's criterion}. 

\vspace{1em}

Near-threshold states are characterized by a large $R_0$. Therefore even a tiny value of $Z$ can give sizable values of $r_0$ and $a$. In particular $r_0$ is expected to be large and negative for $Z\neq 0$.
In the case of the shallow molecular bound state, $Z=0$, we have $a=R_0$ and $r_0=0$. 

\section{On the sign of $r_0$}\label{r0sign}

Weinberg's analysis, as reported in Section~\ref{zinqm}, was originally conceived having in mind the possibility of an elementary deuteron with $Z\neq 0$. The effective range, $r_0$, in that case is referred to the one extracted from low energy $np$ scattering, as due solely to short range interactions. Pion interactions might be included, giving a correction to~\eqref{err0}. Considering that the mass difference $m_n-m_p\ll m_\pi$, the pion plays the role of a heavy degree of freedom and can thus be `integrated out'. We thus expect that, in place of~\eqref{err0}, we should rather have 
\be
r_0=-\frac{Z}{1-Z}R_0+O\left(\frac{1}{m_\pi}\right) \,,
\label{r0_Weinberg}
\ee
as pointed out in Weinberg's paper. 
Finding experimentally an $|r_0|\sim 1/m_\pi\sim 1$~fm would be compatible with $Z=0$: the deuteron is what it is supposed to be, a composite nucleus of a neutron and a proton. If $r_0$ is negative, and has a magnitude sizably larger than 1~fm, that is the token of the presence of an elementary deuteron (and the nuclear deuteron might even be optional). In the case of $np$ scattering it is found that $r_0^{\rm exp}\simeq +1.7$~fm \cite{Bethe, np_r0}, which is considered in line with $Z=0$, assuming that the $1/m_\pi$ correction is positive. We want to point out that $r_0^{\text{exp}}$ is \textit{model-independent}, as the cross section is fitted using the universal expression for the low-energy scattering amplitude \eqref{err0bethe}. On the contrary, as we will comment in Section \ref{purelycompact}, since LHCb is using a Flatté parametrization, it is implicitly assuming that the $X$ is an elementary particle and that no bound state exists.

\vspace{1em}
 
According to an observation made by Landau and Smorodinsky~\cite{Smorod,lqm3-2}, shallow bound states ($Z=0$) with purely attractive potential always have $r_0>0$: this fixes the sign of the $1/m_\pi$ correction, left arbitrary in Weinberg's analysis. 
 
We report here a derivation of this result as provided by Bethe \cite{Bethe}. 
Consider the Schr\"odinger's equation for the reduced radial wave function of the molecular constituents~\cite{Esposito:2021vhu},
\begin{align} \label{pot}
\chi_k^{\prime\prime}(r) +\big[k^2-U(r)\big]\chi_k(r)=0 \,,
\end{align}
with $U(r) \equiv 2 \mu V(r)$ being the potential, which is assumed to be attractive everywhere: $V(r) < 0$. 
We consider the wave function for two values of the momentum: $\chi_{k_{1,2}}\equiv \chi_{1,2}$. A simple manipulation leads to the identity
\begin{align} \label{eqrad}
\chi_2\chi_1^\prime - \chi_2^\prime \chi_1 \Big |_0^R=(k_2^2- k_1^2)\int_0^R dr \, \chi_2 \chi_1 \,,
\end{align}
with $R$ fixed and much larger than the range of the potential, $R\gg 1/m_\pi$.

Consider now the free equation, whose solutions we denote by $\psi$,
\be \psi_k^{\prime\prime}(r) +k^2\psi_k(r)=0 \,, \label{eqm2}\ee 
from which we also obtain
\begin{align} \label{eqpsi}
\psi_2\psi_1^\prime - \psi_2^\prime \psi_1 \Big |_0^R=(k_2^2- k_1^2)\int_0^R dr \, \psi_2 \psi_1\,.
\end{align}
Normalizing the wavefunction such that it is unity at $r=0$, the general expression for $\psi_k$ is
\begin{align}
\psi_k(r) = \frac{\sin(kr+\delta(k))}{\sin\delta(k)}\,,
\end{align}
so that \be\psi^\prime_k(0) = k \cot \delta(k) \,.\ee
The reduced radial wave function $\chi_k$ vanishes at $r=0$, $\chi_k(0)=0$, and we normalize it so that $\chi_k(R)\to \psi_k(R)$ for large enough values of $R$. 

With this proviso, we subtract~\eqref{eqrad} from~\eqref{eqpsi}, to obtain
\begin{align} 
\begin{split} \label{phase2}
&k_2\cot \delta(k_2)-k_1\cot \delta(k_1) =\\ 
&\qquad =(k_2^2- k_1^2)\int_0^\infty dr \, (\psi_2 \psi_1 - \chi_2 \chi_1) \,.
\end{split}
\end{align}
We have extended the integral to infinity, given that it is now convergent due to the same asymptotic behavior of $\psi_k$ and $\chi_k$.

We are now ready to compare the result with the parameters of the scattering amplitude~\eqref{err0bethe}. First we set $k_1=0$ which, recalling that $\lim_{k_1\to0} k_1 \cot \delta(k_1) = -\varkappa_0$, gives, with obvious notation for $k_1=0$,
\begin{align}
k_2 \cot\delta(k_2) = - \varkappa_0 + k_2^2 \int_0^\infty dr\, \left(\psi_2 \psi_0 - \chi_2 \chi_0\right)\,.
\end{align}
Finally, for a shallow bound state, one can further expand for small momenta, $k_2 \cot \delta(k_2) = - \varkappa_0 + \frac{1}{2}r_0 k_2^2 + \dots$, thus finding the formula for $r_0$ 
\begin{align}
r_0=2\int_0^\infty (\psi _0^2-\chi_0^2)\, dr \,.
\label{err0scw}
\end{align}
Let us now introduce $\Delta(r) \equiv \psi_0(r) - \chi_0(r)$, so that $\Delta(0)=1$ and $\Delta(\infty) = 0$.
Subtracting~\eqref{pot} from~\eqref{eqm2} we get\be \Delta^{\prime\prime}(r) = -U(r)\chi_0(r)\,,\ee where $-U(r)>0$ for a purely attractive potential. For the lowest lying bound state (such as the shallow one), $\chi_0(r)$ does not have nodes and can thus be taken to be strictly positive. We thus get $\psi_0''(r) > \chi_0''(r)$ everywhere. This means that (inequalities can be integrated as the continuous sum preserves the inequality)
\be
\int_r^R \psi_0''(r^\prime) \, dr^\prime> \int_r^R \chi_0''(r^\prime)\, dr^\prime \,,
\ee
giving
\be
\psi_0^\prime (R)-\psi_0^\prime (r)> \chi_0^\prime(R)-\chi_0^\prime(r) \,,
\ee
or, for $R\to \infty$, 
\be
\psi_0^\prime (r)< \chi_0^\prime(r) \,.
\ee
Repeating the same steps again we find 
\be
\psi_0 (r)> \chi_0(r) \,.
\label{magg}
\ee
Hence,~\eqref{err0scw} proves that in these conditions $r_0>0$.
If the potential $V$ had a repulsive core we could not conclude that~\eqref{magg} holds for all values of $r$ and therefore $r_0$ itself could be negative, which means that the $1/m_\pi$ correction could be negative. Thus an $r_0\simeq -1$~fm experimental value would not necessarily be compatible with $Z=0$, in this case, but an $r_0=-5$~fm, for example, should anyway be ascribed to a $Z>0$ value. 

The case of the $X(3872)$ is more complicated than that of the deuteron, in this respects. As we will see in Section~\ref{corrs}, the corrections due to pion interaction could in principle be larger, because they are expected to be of order $1/\mu$, with $\mu \ll m_\pi$. But this is not what will happen, as it is shown in Section~\ref{correpion}.

\section{The potential generating the shallow bound state}
\label{corrs}
The unspecified potential $V$ responsible for the shallow bound state, as the one treated in Section~\ref{zinqm}, has typically a range $R$ which is way smaller than the extension of the universal $S$-wave wavefunction, which we can derive from~\eqref{universalreduced}
\be
\Psi(r)=Y_{00}\,\frac{\chi(r)}{r}=\left(\frac{2mB}{4\pi^2}\right)^{1/4}\frac{\exp\big(-r\sqrt{2mB}\big)}{r} \,.
\label{universal}
\ee 
Since $\Psi(r)$ extends over a broad range, $1/\sqrt{2mB} \gg R$, and since its form does not depend on the details of the potential itself, one can model the interaction via the attractive Dirac delta function potential, $V_s(r)=-\lambda_s\delta^3 (\bm r)$, which is the non-relativistic limit of the effective $(D\bar D^*)^2$ interaction. As shown in~\cite{jackiw}, this potential is irregular, and leads to UV divergences, which need to be renormalized away. This is done reabsorbing them in a renormalized coupling, $\lambda \propto 1/\sqrt{B}$, with $B$ the binding energy of the only allowed bound state. In this framework, the smalleness of the binding energy arises from a microscopic tuning between the bare coupling, $\lambda_s$, and the UV cutoff.\footnote{Consider the regularized potential $$V=\frac{\lambda}{4\pi R^2}\delta(r-R)\qquad \text{for}\qquad R\to 0$$ in place of $V=\lambda\delta^3(r)$. It can be shown that the relation with the binding energy is $$\frac{1}{\lambda m}=\frac{1}{2\pi R}-\frac{\sqrt{2m B}}{2\pi}$$ where $m$ is some reduced hadron mass and $R\sim 3$~fm, the range of strong interactions. Since $1/(\lambda m)\ll 1/(2\pi R), \, \sqrt{2mB}/2\pi$  we get $$B\sim (197/3)^2/(2 m)\sim 4~\text{MeV}$$ to be compared with $B=2.2$~MeV for the deuteron.} (For a quantum field theory treatment see, e.g.,~\cite{Braaten:2003he}.)

Incidentally, notice also that the expectation value of the kinetic energy on $\Psi$ is
\be
\langle {\rm K.E.}\rangle_{\Psi}=\int_0^\infty \Psi^*(r) \frac{-1}{2 m r^2}\frac{\partial }{\partial r}\left(r^2\frac{\partial}{\partial r}\Psi(r)\right) \, 4\pi r^2 dr=-B \,,
\ee
which also means that the expectation value of the momentum is\footnote{In the case of a Dirac delta potential $V(r)=\lambda \,\frac{\delta(r)}{4\pi r^2}\Rightarrow \langle V\rangle_{\Psi}=0$ so that $-B=\langle H\rangle=\langle T\rangle+\langle V\rangle=\langle T\rangle=\frac{\langle p^2\rangle}{2m} $. This corresponds to the pole in the $E+B$ denominator or $\langle p^2\rangle_{\Psi}=-2 m B$.}
\be
\sqrt{|\langle p^2\rangle_\Psi|} = \sqrt{2 m B}~(\sim 1/\Delta x)\simeq 14~{\rm MeV} \,,
\label{virial}
\ee
for a binding energy of 100~keV. This is the typical value expected for the relative momentum in the center-of-mass of a loosely bound molecule. As discussed in~\cite{prod1,prod2,prod3,prod4}, this momentum is too small to explain the large cross section for formation of such a loosely bound molecule in prompt $pp(\bar p)$ collisions. From here the hypothesis of partial compositeness of the $X$: it is the elementary core of the $X$ to allow prompt production.

The unperturbed wave function of the $V_s$ potential bound state is computed exactly in\cite{jackiw} and it corresponds to that in~\eqref{universal}. A weak perturbation to the strong potential $V_s(r)$ derives from the weak potential $V_w(r)$ due to one-pion exchange in the $u$-exchange $D\bar D^*$ channel. 
This turns out to be a complex potential, as discussed in\cite{Esposito:2023mxw}, of the form 
\be
V_w(r)=-\alpha\frac{e^{i\mu r}}{r} \,,
\label{weak}
\ee
where $\alpha$ weights the weak pion interactions
\be
\alpha=\frac{g^2\mu^2}{24\pi f_\pi^2} \simeq 5\times 10^{-4} \,, \qquad \text{ with } \qquad f_\pi\simeq 132~\text{MeV} \,,
\ee
and 
\be
\mu=\sqrt{2m_\pi \delta}\simeq 43~\text{MeV} \,, \qquad \text{ with } \qquad \delta=m_{D^*}-m_{D}-m_\pi \simeq 7~\text{MeV} \,.
\label{muscale}
\ee
This potential is a consequence of the fact that, in the non-relativistic limit, the pion exchange in the $u$-channel is described by an amplitude proportional to 
\be
\int \frac{q_iq_j \, e^{i\mathbf q\cdot \mathbf r}}{q^2+m_\pi^2-i\epsilon}\,\frac{d^3q}{(2\pi)^3}
\underset{\;\text{NR}\;}{\longrightarrow}
\int \frac{q_iq_j \, e^{i\mathbf q\cdot \mathbf r}}{\bm q^2-\mu^2-i\epsilon}\,\frac{d^3q}{(2\pi)^3} \,,
\ee
with the momenta in the numerator arising from the derivative couplings of the pion. If we neglect the the mass $\mu$ in the previous formula we find that 
\be
\int \frac{q_iq_j \, e^{i\mathbf q\cdot \mathbf r}}{\mathbf q^2{\color{yellow}}-i\epsilon}\,\frac{d^3q}{(2\pi)^3} =-\frac{1}{4\pi} \left(\frac{3\hat{r}_i\hat{r}_j}{r^3}-\frac{\delta_{ij}}{r^3}-\frac{4\pi}{3}\delta^3 (\boldsymbol r)\right) \,.
\ee
The $1/r^3$ potential does not allow bound states (use the $S$-wave average $\langle \hat{r}_i\hat{r}_j\rangle=\frac{1}{3}\delta_{ij}$). The $\delta^3(r)$ potential
can have bound states, upon a renormalization of the coupling, even though the actual binding energy $B$ is not calculable.

Keeping $\mu$ finite at its actual value we have 
\be
V_w=-\frac{g^2}{2f_\pi^2}\int \frac{q_iq_j \, e^{i\mathbf q\cdot \mathbf r}}{\mathbf q^2-\mu^2-i\epsilon}\,\frac{d^3q}{(2\pi)^3}=-\underbrace{\frac{g^2}{6f_\pi^2}}_{\beta}\left(\delta^3(r)+\mu^2\frac{e^{i\mu r}}{4\pi r}\right)\delta_{ij} \,,
\ee
so that the total potential is
\be
V\equiv V_s+V_w=-\left(\lambda_s+ \beta\right)\delta^3(\boldsymbol r)- \alpha\frac{e^{i\mu r}}{r} \,,
\ee
as in~\eqref{weak}.

This analysis shows that pion interactions generated by $V_w$ could generate corrections to $r_0$ and $a$, determined in formulae~\eqref{err0} and~\eqref{aa}, of order $1/\mu$, with no clear indication on the sign. 
However $1/\mu\sim 5$~fm. If $Z=0$ and the correction due to one-pion interaction is indeed negative and as sizeable as $-5$~fm, we would have a negative and large $r_0$ even with $Z=0$. This could make the application of Weinberg's argument questionable.

\section{Corrections to $r_0$ due to one-pion exchange}
\label{correpion}
The correction to the scattering amplitude for a purely molecular state, due to one-pion exchange, can be written as
\be
f=f_s + f_w =\frac{1}{-\frac{1}{a}-ik} +f_w \,,
\ee
where in the first term we take $r_0=0$, as $Z=0$ for the molecule. 

The correction is calculated through the formula for the so-called distorted wave Born approximation~\cite{Esposito:2023mxw,DWBA},
\be
f_w=-\frac{2m}{4k^2} \int V_w(r) \,\chi_s^2(r) \, d r \,,
\ee
where $\chi_s$ are the scattering reduced wavefunctions in the $\delta^3(r)$ potential, and $m$ is the reduced mass of the $D\bar D^*$ system. 
Therefore $r_0$ due to pion exchange is found as the $k^2$ coefficient in the expansion around $k=0$ and $\alpha=0$ of 
\be
f^{-1}=\left(\frac{1}{-\frac{1}{a} - i k}-\frac{2m}{4k^2} \int V_w(r) \,\chi_s^2(r) \, d r\right)^{-1} \,.
\label{f_inv}
\ee

The expression for $f_w$ can be derived with the following argument. Let's consider the Born formula for the scattering amplitude 
\be
f_{\rm Born}=-\frac{m}{2\pi}\int V(r)\, e^{i(\boldsymbol k-\boldsymbol k^\prime)\cdot \boldsymbol r} \, d^3r \,,
\ee
and expand
\be
e^{i\boldsymbol k\cdot \boldsymbol r}=\sum_{\ell=0}^{\infty} i^\ell j_\ell(k r)(2\ell+1)P_\ell(\boldsymbol {\hat k}\cdot \boldsymbol {\hat r}) \,,
\ee
and 
\be
e^{-i\boldsymbol k^\prime \cdot \boldsymbol r}=\sum_{\ell=0}^{\infty} i^\ell j_\ell(k^\prime r)(2\ell+1)(-1)^\ell P_\ell(\boldsymbol {\hat k^\prime}\cdot \boldsymbol {\hat r}) \,.
\ee
Using the result
\be
\int P_\ell (\boldsymbol n_1 \cdot \boldsymbol n_2) P_{\ell^\prime}(\boldsymbol n_1\cdot \boldsymbol n_3)d\Omega_1=\delta_{\ell \ell^\prime}\frac{4\pi}{(2\ell+1)}P_\ell(\boldsymbol n_2\cdot \boldsymbol n_3) \,,
\ee
and the fact that $(-1)^\ell i^{2\ell}=+1$ for every $\ell$, and $k=k^\prime$ for elastic collisions, we get 
\be
f=-2m \sum_{\ell=0}^{\infty}(2\ell+1)P_\ell(\cos\theta) \int V(r) (j_\ell(k r))^2 r^2 d r \,,
\ee
to be compared with the standard formula
\be
f=\sum_{\ell=0}^{\infty}(2\ell+1)P_{\ell}(\cos\theta)\frac{e^{i\delta_{\ell}}\sin\delta_{\delta}}{k} \,.
\ee
The comparison gives
\be
f_{\ell}=\frac{e^{i\delta_\ell}\sin\delta_\ell}{k}=-2m \int V(r) (j_\ell(k r))^2 r^2 d r \,.
\ee
Setting 
\be
\chi_\ell^{(0)}(r)=2kr \, j_\ell(kr) \,,
\ee
we write the general expression
\be
f_{\ell}=-\frac{2m}{4k^2} \int V(r)\, \big(\chi_\ell^{(0)}(r)\big)^2\, d r \,.
\ee
The distorted wave Born Approximation consists in replacing the free reduced wavefunctions $\chi^{(0)}$ with the scattering wavefunctions, $\chi_s$, obtained from the the strong potential, $V_s$, which here corresponds to the Dirac-delta potential
\be
f_w=-\frac{2m}{4k^2} \int_0^\infty V_w(r)\, \big(\chi_s(r)\big)^2\, d r \,.
\label{eq:f_w_9.14}
\ee 

The explicit calculation of $f_w$ has to be done in a regularization scheme, since the integral does not converge at short distances. A practical way to do the calculation in the case at hand is the following. Use $\exp(-\mu r)$ in place of $\exp (i\mu r)$ in the expression of $V_w$. In the final result send $\mu\to -i\mu$. 
Introduce the regularized 
\be
\chi_{s}^{\rm I}(r)=2 k r\left(\frac{e^{i\delta}\sin(k r+\delta)}{k r}-\frac{e^{i\delta}\sin \delta}{k r}\right) \,,
\ee
for $r\in [0,\lambda]$, and 
\be
\chi_s^{\rm II}(r)=2kr\left(\frac{e^{i\delta}\sin(k r+\delta)}{k r}\right) \,,
\ee
for $r\in [\lambda,\infty)$. The latter is the scattering function as given in~\cite{jackiw}. The integral obtained this way is finite. Now, use $\delta=\cot^{-1}\left(-1/k a\right)$, expand the result to second order around $k=0$, and to first order around $\alpha=0$, and finally take the limit $\lambda\to 0$. In the final result set $\mu\to -i\mu$ to obtain
\be
r_0=2m \alpha \left( \frac{2}{\mu ^2 a^2}+\frac{8i}{3 \mu a}-1\right) \,.
\ee
Numerically this gives~\cite{Esposito:2023mxw}
\be
-0.20 \text{ fm} \lesssim \text{ Re} \,r_0 \lesssim - 0.15 \text{ fm} \,,
\label{eq:r_o_esposito}
\ee
and
\be
0 \text{ fm} \lesssim \text{ Im} \, r_0 \lesssim 0.17 \text{ fm} \,.
\ee
The result found agrees analytically to what found by Braaten \cite{Braaten:2020nmc}. The corrections to $r_0$ due to pion exchange are indeed {\it negative}: they push $r_0$ to negative values, thus mimicking a positive $Z>0$, even for a pure bound state. However they are very small in size, as a consequence of the pion's weak coupling: a value of $r_0\simeq -5$~fm, for example, cannot be due to one-pion corrections. 

This tells us that $r_0$ is a potentially very good parameter to discriminate the nature of a particle like the $X(3872)$, in a model independent way.

\section{The structure of the $X$ from its lineshape.}\label{zfls}

The discrete level of the molecular $X$, at $E=-B$, is actually quasi-discrete, since it is broadened by the width, $\Gamma$, due to various decay channels of the $X$. The discussion carried on so far refers strictly to the scattering of two stable particles that resonate on a shallow, stable, bound state. Nonetheless, it is known that the $X$ couples to other channels too, as $J/\psi \pi\pi$, although we assume, as supported by data, that a single channel dominates \cite{ParticleDataGroup:2024cfk}. 

In the following we will remind some few facts about the Breit-Wigner and the Flatt\'e functions~\cite{Flatte}. In particular the latter has been used to extract from data the value of $Z$ associated to the $X$ field~\cite{LHCb:2020xds}.
As we discuss below, this approach is actually only testing the compact nature of the $X(3872)$. A good fit with the Flatt\'e lineshape means that we have a good test of the elementariness of the $X(3872)$, but it does not tell anything about the bound state hypothesis. In Section~\ref{flatt} we give a more exhaustive derivation of the Flatt\'e scattering amplitude based on effective field theory. 

\vspace{1em}
 
{\bf \emph{ Breit-Wigner.}} In the non-relativistic quantum mechanics description, when a particle decays, the boundary conditions on the wavefunctions at infinite distance require an outgoing spherical wave. This complex boundary condition means that the energy eigenvalues themselves can be complex: $E$ can have an imaginary part. The time evolution factor of a quasi-stationary state is $\exp(-iEt)$. If $E=E_0-i\Gamma/2$, the probability of finding the particle not-decayed at time $t$ is $\exp(-\Gamma t)$. Again we require no-incoming spherical wave in the asymptotic conditions, therefore, just as we did in~\eqref{noing} (for the $\Gamma=0$ case), we have to require that 
\be
\mathcal{B}\left(E_0-i\frac{\Gamma}{2}\right)=0 \,.
\label{zeroB}
\ee
Let us take again $E_0>0$, as done in Section~\ref{SA}, but this time expanding $\mathcal{B}(E)$ around $E_0$, and not around $-E_0$, as it was done in~\eqref{Bexpand}. Recall that in Section~\ref{SA} we assumed $E_0=B=$~binding energy. So this time we are not starting from the analysis of a bound state: we are studying a scattering problem, i.e. above threshold, and we mean to describe a standard resonance, i.e. a compact bound state of quarks. 

The energy $E$ is on the positive part of the real axis, and in addition it is shifted on the lower half of the complex plane, since $\Gamma>0$. The complex plane $z=\sqrt{-E}$ is cut for real positive values. The point $E_0$ on the upper edge of the cut can be brought in the lower half plane by going around the branch point at $z=0$, along some path $\gamma$, which produces $\Delta_\gamma\arg z=(2\pi-0)/2=\pi$. This would change the outgoing spherical wave $\exp(ikr)$ in the incoming one and viceversa. To avoid this (we want to suppress the incoming spherical wave), one has to choose a path $\gamma$ going simply through the cut, on the second Riemann sheet, rather than going around $z=0$. 

Let us consider positive energy values close to the quasi-discrete level $E_0$, assuming that $\Gamma$ is small, which is certainly the case for the $X$. We consider now the expansion of $\mathcal{B}(E)$ in terms of the difference $E-(E_0-i\Gamma/2)$ which, using~\eqref{zeroB}, at the first order gives 
\be
\mathcal{B}(E)\simeq\beta (E-E_0+i\Gamma/2) \,.
\ee
where $\beta$ is some complex constant --- to be compared with~\eqref{sette}. 
As in footnote~\ref{foot1} we have $\mathcal{A}(E)=\mathcal{B}^*(E)$, therefore 
\be
\chi(r)=\beta^* (E-E_0-i\Gamma/2)\exp(ikr)+\beta (E-E_0+i\Gamma/2)\exp(-ikr) \,,
\ee
so that, as done in~\eqref{nove}, we get the $S$-wave phase shift of this function\footnote{An extra factor of $(-1)^\ell$ should be included in presence of $\ell\neq 0$.}
\be
\exp(2i\delta_0)=
\exp(2i\delta_0^0)\left(1-\frac{i\Gamma}{E-E_0+i\Gamma/2}\right) \,,
\label{Sres}
\ee
where $\exp(2i\delta_0^0)\equiv -\beta^*/\beta$ and $\delta_0^0$ is clearly the phase $\delta_0$ far from resonance: $E-E_0$ large with respect to $\Gamma$. 
Once the result~\eqref{Sres} is plugged into the general formula
\be
f(\theta)=\frac{1}{2ik}\sum_{\ell\geq 0}(2\ell+1)(\exp(2i\delta_\ell)-1)P_\ell(\cos\theta) \,,
\ee
and we consider the scattering of slow particles, $k\to 0$, where only $S$-wave is important,\footnote{\label{theo} From the $f_\ell\sim\delta_\ell/k\sim k^{2\ell}$ behavior at low energies. Indeed analyticity requires $M$ (first line of~\eqref{deri}) to scale as $k^{\ell+1/2}k^{\prime\ell^\prime +1/2}$ for $k\to 0$, and $f$ in the center-of-mass contains $1/k$, normalizing the states appropriately (see Weinberg-I~(3.7.5) \cite{Weinberg_QFT}). So if $\ell=\ell^\prime$ we have $f\sim k^{2\ell}$ at low energies. } we get
\be
f=-a-\frac{1}{k}\frac{\Gamma/2}{E-E_0+i\Gamma/2} \,,
\label{bw}
\ee 
since\footnote{ \label{footx} At small $k$ one finds from the asymptotic form of the radial wavefunction at $kr\gg 1$ that $\delta_\ell\simeq k^{2\ell+1}\frac{c_2}{c_1} (2\ell-1)!!(2\ell+1)!!$. The ratio of coefficients $c_2/c_1$ is defined as $-a$ (the scattering length). } $\exp(2i\delta_0^0)\simeq 1$ for $\delta_0^0=-a k\ll1$. The scattering length $-a$ term in $f$ is due to the scattering amplitude far from resonance, i.e. the simple potential scattering amplitude for $k\to 0$, independently of the quasi-stationary state. In the vicinity of the resonance (and for $\Gamma\ll E_0$) the second term dominates and we can consider
\be
f=-\frac{1}{k}\frac{\Gamma/2}{E-E_0+i\Gamma/2} \,.
\label{tbw}
\ee 

\vspace{1em}

{\bf \emph{The case $\bm{E_0\to 0}$.}} Now consider the case in which $E_0$ is almost vanishing --- i.e. closer to zero than to the next level, if any --- and consider $E\sim E_0\sim 0$. We see that in this limit the resonant term in $f$ has a $f\sim 1/k$ behavior, contrarily to the general result according to which at low energy $f\sim\text{constant}$, as reminded briefly in footnote~\ref{theo}. This means that~\eqref{tbw} is not adequate to describe the case in which $E\sim E_0\sim 0$. On the other hand this is the case of interest to our discussion, since the quasi-stationary level of the $X$ is found very close to the $D\bar D^*$ threshold, which can be taken as the offset of the energy values ($E_0\sim 0$ from above). In this sense, formula~\eqref{tbw} describes an (ordinary) resonance, {\it above} threshold. What we will find now is that, in the $E_0\to 0$ case, the $1/k=1/\sqrt{2mE}$ in~\eqref{tbw} has to be substituted by $1/k\to 1/\sqrt{2mE_0}$. 

The issue is in that the $E=0$ value is a branch point of $\mathcal{B}(E)$ and in going around it from the upper to the lower edge of the cut, something that we have to do to introduce a finite width, changes $\mathcal{B}(E)\to \mathcal{B}(E^*)= \mathcal{B}^*(E)$. If the expansion of $\mathcal{B}(E)$ is in powers of $z=\sqrt{-E}$, in going around $E=0$, we change sign to $z$, which is equivalent to changing $\mathcal{B}(E)$ into $\mathcal{B}^*(E)$ in the following expression where we introduce the real constants $\epsilon_0$ and $\gamma$  
\be
\mathcal{B}(E)=\beta(E)(E-\epsilon_0+i\gamma\sqrt{E}) \,,
\ee 
where indeed we have $i\sqrt{E}$ on the upper edge of the cut, and $-i\sqrt{E}$ on the lower edge of the cut, and we assume $\epsilon_0>0,\gamma>0$. The function $\beta(E)$ is not expected to have a zero at $E=0$. As we did in~\eqref{zeroB} we have to require $(E-\epsilon_0+i\gamma\sqrt{E})$ to go to zero for $E\to E_0-i\Gamma/2$. This translates into
\be
(E_0-i\Gamma/2)-\epsilon_0+i\gamma\sqrt{E_0-i\Gamma/2}=0 \,.
\label{conditio}
\ee
For small $\Gamma\ll E_0$ we get 
\be
\sqrt{E_0-i\Gamma/2}=\sqrt{E_0}-\frac{i \Gamma }{4 \sqrt{E_0}}+O\left(\Gamma ^2\right) \,.
\ee
Substituting $E_0=\epsilon_0$ in~\eqref{conditio} we get
\be
-i\Gamma/2+i\gamma\sqrt{\epsilon_0}+O(\gamma\Gamma)=0 \,,
\ee 
which is zero, at first order in $\gamma$, if
\be
\gamma=\frac{\Gamma}{2\sqrt{\epsilon_0}} \,.
\ee
Therefore we must substitute $E_0\to \epsilon_0$ and $\Gamma\to 2\gamma\sqrt{E_0}$ in the second term in~\eqref{bw}, which describes $E$ close to $E_0(=\epsilon_0)$, 
to get
\be
f=-a-\frac{1}{\sqrt{2m}}\frac{\gamma}{(E-E_0+i\gamma\sqrt{E})} \,.
\ee
In the vicinity of $E_0$ --- closer to zero than to the next level --- we can write therefore
\be
f=-\frac{1}{\sqrt{2mE_0}}\frac{\Gamma/2}{(E-E_0+i(\Gamma/2)\frac{\sqrt{2m E}}{\sqrt{2mE_0}})}\simeq -\frac{1}{\sqrt{2mE_0}}\frac{\Gamma/2}{(E-E_0+i\Gamma/2)} \,,
\label{prec}
\ee
for $E\simeq E_0$ which is like the formula in~\eqref{tbw}, but with $1/k\to 1/\sqrt{2mE_0}$. 
If in the first term on the right hand side of Eq.~\eqref{prec} we fix $E_0=m_{\rm BW}>0$ and $\Gamma\simeq g^2\sqrt{2m m_{\rm BW}}$, using the non-relativistic formula for the decay momentum, we obtain 
\be
f= -\frac{g^2/2}{E-m_{\rm BW}+i\, g^2/2 \, \sqrt{2m E}} \quad \text{ with } \quad m_{\rm BW} > 0 \,.
\label{tf}
\ee
in place of~\eqref{tbw}. We remind here that this is a Breit-Wigner formula for a resonance slightly above threshold, which decays non-relativistically in its components whose reduced mass is $m$. 

\vspace{1em}

{\bf {\emph{Single channel Flatt\'e continuation.}}} If we change the sign to $E_0$, using $E_0\lesssim 0$, because of the cancellation observed above between $\Gamma$ and $\sqrt{2mE_0}$ we simply get from the first term on the right hand side of~\eqref{prec}
\be
f= -\frac{g^2/2}{E-m_{\rm F}+i\, g^2/2 \, \sqrt{2m E}}\quad \text{ with } \quad m_{\rm F}<0 \,.
\label{tf2}
\ee
This means that we are simply continuing to the case in which the elementary resonance is found slightly below threshold, rather than above. In the previous formula let us assume that 
\be
E-m_{\rm F}+i\, \frac{g^2}{2} \, \sqrt{2m E}=0\qquad \text{for}\qquad E=-B \,.
\ee
In the complex variable $z=\sqrt{E}$, the value of $E$ should be changed into $-B$ going through a path $\gamma$ in the second unphysical sheet, so that $\Delta_\gamma \arg z=(-\pi-0)/2$ and 
\be
\sqrt{E}\big|_{E>0} \mapsto \exp(-i\pi/2)\sqrt{B}=-i \sqrt{B}\big|_{B>0} \,.
\ee
This gives 
\be
-m_F=B-g^2/2\sqrt{2mB} \,,
\ee
which we plug back into~\eqref{tf2}. Expanding the term $(\sqrt{2mB}-i\sqrt{2mE})$, with $E$ around $B=0$, we get 
\be
f=-\frac{g^2/2}{1-g^2/2\sqrt{m/2B}}\frac{1}{E+B}\simeq -\sqrt{\frac{2B}{m}}\frac{1}{E+B} \,,
\ee
which holds at {\it low energy} and for a very low value of the distance from threshold $B$ | here the meaning of $B$, as clarified above, is not that of a bound state energy (binding energies are on the physical sheet). 
On the other hand it is clear that at low energy and small $B$ this is the same parametrization as in~\eqref{main}, albeit with a different meaning of $B$. When comparing data on the lineshape of $X$ with the Flatt\'e parametrization a broader region of energies is included, beyond $E\simeq -B\simeq 0$. The resulting fit will give a value for the coupling $g$ of the (elementary) resonance, and the parameter $m_F$. We observe that this coupling is related to $g_c$ defined in~\eqref{418}. 

As we will see in Section~\ref{purelycompact}, consistently with what observed here, the Flatt\'e parametrization can be deduced by a non-relativistic effective theory in which only the trilinear coupling $\mathfrak XD\bar D^*$ is included and therefore there is no possibility of forming $D\bar D^*$ bound states. In Section~\ref{purelycompact}, differently from what has been done here, the most general coupled channel case is considered. The coupled channel formula is the one which has indeed been used by the LHCb collaboration in the analysis described in the next paragraphs. 

\vspace{1em}

{\bf {\emph{The LHCb analysis. }}}The expression~\eqref{tf2} is similar to that used by the LHCb collaboration for the determination of $f$ from data on the lineshape of the $X$~\cite{LHCb:2020xds,Esposito:2021vhu} in that they also treat, and determine experimentally, the case $E_0<0$ ($E_0=m_F<0$)
\be
f= -\frac{N}{E-m_F+i(g_{\rm LHCb}/2)\Big(\sqrt{2m E}+\sqrt{2m_{+} (E-\Delta)}\Big)} \,.
\label{tolhcb}
\ee
Differently from what obtained above, we have two square root terms in the denominator,\footnote{An analysis using the full width of the $D^*$ was proposed in \cite{mikha} for the $T_{cc}^+$. This approach could potentially be extended to the case of the $X(3872)$.} referring to the neutral and charged open charm meson thresholds --- we will give a derivation of this (coupled channel) formula in Section~\ref{flatt}. For the moment we can take this part for granted and simply comment on what was done in the LHCb analysis. In the LHCb formula $m\simeq 967~\text{MeV}$ is the reduced mass of the neutral $D\bar D^*$ pair as defined above, whereas $m_+\simeq 969~\text{MeV}$ is the reduced mass of the charged pair. The parameter $\Delta$ is given by the difference between the charged and neutral thresholds 
\be
\Delta =\Delta_+-\Delta_0= (m_{D^{*-}}+m_{D^+})-(m_{D^{*0}}+m_{D^0}) \simeq 8.2~\text{MeV} \,,
\label{lhcb}
\ee
whereas $E$ is the $J/\psi\pi\pi$ invariant mass, $M=\sqrt{-(p_J+p_\pi+p_\pi)^2}$, in the $X\to J/\psi\pi\pi$ decay as measured from the offset at $\Delta_0$
\be
E=M- \Delta_0 \,.
\ee
This way, $E-\Delta$ in~\eqref{lhcb} corresponds to 
\be 
E-\Delta=M-\Delta_+ \,.
\ee 
Therefore we can write
\be
f= -\frac{N}{(M-\Delta_0)-m_F+i(g_{\rm LHCb}/2)\Big(\sqrt{2m (M-\Delta_0)}+\sqrt{2m_{+} (M-\Delta_+)}\Big)} \,,
\ee
where $m_F$ itself is determined with respect to the threshold (to the offset, as in~\eqref{tolhcb}): in place of $m_F$ we should write $m_F^0-\Delta_0$ so that 
\be
f= -\frac{N}{M-m_F^0+i(g_{\rm LHCb}/2)\Big(\sqrt{2m (M-\Delta_0)}+\sqrt{2m_{+} (M-\Delta_+)}\Big)} \,.
\label{finlhc}
\ee
The square root of the energy terms in~\eqref{tolhcb}, derived in the analysis done above, can be obtained by $\Gamma$ in~\eqref{f11} using the methods of quantum field theory (independently from the sign of $E_0$) where $\Gamma$ is connected to the imaginary part of the bubble diagrams in the propagator of the $X$. As we will see, the imaginary part of the loop integral describing 
$X\to D\bar D^*\to X$ is proportional to $\sqrt{2mE}$ where $E$ is referred to the threshold $\Delta_0$ or $\Delta_+$ depending on the fact that the particles running in the loop are $D^0\bar D^{*0}$ or $D^+\bar D^{*-}$; therefore in~\eqref{finlhc} both contributions are included (in the derivation reported above we have seen how these terms arise in the non-relativistic quantum mechanics formalism). 

LHCb fit to data determines 
\be 
m_F\simeq -7.18~\text{MeV},\quad g_{\text{LHCb}}=0.108\pm0.003 \,.  \label{gLHCbfit}
\ee
These values are obtained by fitting the data with the expression \eqref{finlhc} for the scattering amplitude on the whole range of $E$ available. 

In order to obtain $a$ and $r_0$, we have to expand~\eqref{finlhc} for small values of $E$ 
\be
f=-\frac{(2N/g_{\rm LHCb})}{(2/g_{\rm LHCb})(E-m_F)-\sqrt{2m_+\Delta}+E\sqrt{m_+/2\Delta}+ik} \,,
\label{ikappa}
\ee
where $k=\sqrt{2mE}$, so that we can compare to the universal expression for the scattering amplitude at low energies
\be
f=\frac{1}{ -\frac{1}{a}+\frac{1}{2}r_0 k^2-ik} \,.
\label{eq:univ}
\ee 
From this we deduce\footnote{As mentioned in footnote \ref{footnote5}, the phase in front of the scattering amplitude is arbitrary \cite{Weinberg_QFT}. To compare \eqref{ikappa} with the universal amplitude \eqref{eq:univ}, we can add a minus sign in front of the latter. This is equivalent to considering an overall minus sign in \eqref{deri} and proceed accordingly.} 
\begin{align}
 \frac{1}{a}&=-\frac{2m_F}{g_{\rm LHCb}}-\sqrt{2m_+\Delta}\simeq 6.92~\text{MeV} \,, \notag\\
r_0&=-\frac{2}{m g_{\rm LHCb}}-\sqrt{\frac{m_+}{2m^2 \Delta}}\simeq -5.34~\text{fm}\,.
\label{eq:a0_LHCb}
\end{align}
This corresponds to a positive value of the scattering length $a\simeq 28$~fm.

\vspace{1em}

In \cite{r0_isospin}, it was noted that the scattering amplitude \eqref{ikappa} accounts for the simultaneous effect of both the neutral channel $\Bar{D}^0{D}^{*0}$ and the charged channel $D^-D^{*+}$ (coupled channel scattering), whereas Weinberg's derivation for $Z$ is formulated for single-channel scattering \cite{Weinberg:1965zz}. It has been proposed that, to compare with Weinberg's results (formulae~\eqref{err0} and~\eqref{aa}) all isospin symmetry-violating terms related to the charged channel must be subtracted from $a$ and $r_0$, before extracting $Z$~\cite{r0_isospin}. For instance, in \eqref{eq:a0_LHCb}, the term $\Delta r^I_F=\sqrt{m_+/(2m^2\Delta)}$, which depends on $\Delta$, should be removed from $r_0$, yielding
\begin{equation}
 r_0^\prime=r_0+\Delta r^I_F\simeq -3.78\,\text{fm} \,.
 \label{recipe}
\end{equation}
Using $r_0^\prime$ to solve~\eqref{err0} and~\eqref{aa} , we get $Z\simeq0.11$ and $ R_0\simeq 30\,\text{fm}$ corresponding to $B\simeq23 \text{ keV}$.\footnote{We cannot find a rigorous proof supporting this recipe.}

However, as it will be shown in the next Section, the Flatt\'e amplitude is derived in the effective field theory by forbidding the molecular bound state $\mathfrak B$ to begin with. The quartic interaction is not included and therefore 
we do not expect any bound state in the effective theory. 
Considering this, we should not compare the results of $a$ and $r_0$ (with or without isospin correction) to Weinberg's formulae depending on $Z$, given that no partial compositeness is possibly included in the Flatt\'e fit. The negative $r_0$ obtained from the fit is not related to Weinberg's formula \eqref{err0}, but it is a property of the Flatté amplitude, as we will demonstrate in Section \ref{purelycompact} (see Eq.~\eqref{eq:as_r0_flatte}).

The coupling $g_{\text{LHCb}}$ should be connected to the phenomenological coupling $g_c$ in \eqref{418} for the compact particle, as we will see in Section \ref{purelycompact}, rather than to the coupling $g_Z$ of the bound state in \eqref{wei2}. A good fit to a Flatt\'e distribution here means a positive test of the elementary particle hypothesis for the $X(3872)$.

The recipe~\eqref{recipe} proposed in \cite{r0_isospin} should be applied, with appropriate modifications, to the purely molecular case discussed in Section \ref{sec:pure_mol}. Regarding this, it has been recently pointed out \cite{Dong_conf} that using a scattering amplitude that considers only mesonic interactions, without the mediation of an elementary $X$, and taking into account both the effect of the coupled-channel and the effect of the pion, leads to a value of $r_0 > 0$ compatible with a purely molecular hypothesis.

\section{The effective theory of the $X$ and $D\bar D^*$ mesons}
\label{flatt}
A low energy effective field theory of pseudoscalar and vector open charm mesons $D^{*0}, \Bar{D}^{0}$ and $D^{*+},D^{-}$ can be formulated, allowing or not a coupling to an elementary\footnote{Elementary in the sense of a compact state of quarks as elementary as the $D$ and $D^*$ mesons are.} \textit{isosinglet} $X$  \cite{Braaten}.

We define the isospin doublets for scalar mesons\footnote{We recall that in the non-relativistic treatment, the fields $D/\bar D$ destroy the particles/anti-particles, while $D^\dagger/\bar D^\dagger$ ones create them.}
\begin{equation}
 D=\begin{pmatrix}
 D^+\\
 D^0
 \end{pmatrix} \,,
 \quad\quad
 \Bar{D}=\begin{pmatrix}
 \Bar{D}^0\\
 D^-
 \end{pmatrix} \,,
\end{equation}
and in a similar way the vector ones, which we denote by $\bm{D}$ (the vector symbol standing for the spin indices). The charge conjugation operator $C$ is represented by the combination of the unitary operator $K$,
\begin{equation}
 K\,D=\begin{pmatrix}
 D^-\\
 \Bar{D}^0
 \end{pmatrix} \,, \quad\quad
 K\,\bm{D}=\begin{pmatrix}
 \bm D^{*-}\\
 \bm{\Bar{D}}^{*0}
 \end{pmatrix} \,,
\end{equation}
and the first Pauli matrix, $\sigma_1$
\begin{equation}
 D\xrightarrow{\;C\;}\Bar{D}=\sigma_1\,K\,D \,,\quad\quad \bm{D}\xrightarrow{\;C\;}\Bar{\bm{D}}=\sigma_1\,K\,\bm{D} \,.
\end{equation}
The $X$ has $C=1^+$, therefore we write the following isospin singlet and triplet positive charge conjugation currents ($\sigma_a$ are the Pauli matrices and $\epsilon=i\sigma_2$)
\begin{align}
 \bm J_S={}&\Bar{D}^{\rm t}\epsilon\,\bm{D}-D^{\rm t}\epsilon\,\Bar{\bm{D}} \,, \\
 \bm J_T^a={}&\Bar{D}^{\rm t}\epsilon\,\sigma_a\,\bm{D}+D^{\rm t}\epsilon\,\sigma_a\,\Bar{\bm{D}} \,,
\end{align}
Here ``t'' stands for the transpose of the isospin doublet. In a more explicit form, these read (for the triplet case we only take the example of the $J_T^3$)
\begin{align}
 \bm J_S={}&\Bar{D}^{0}\bm{D}^0+D^{0}\Bar{\bm{D}}^0-\left(D^{-}\bm{D}^+ + D^{+}\bm{D}^-\right) \,,\\
 \bm J_T^3={}&-\Bar{D}^{0}\bm{D}^0-D^{0}\Bar{\bm{D}}^0-D^{-}\bm{D}^+-D^{+}\bm{D}^- \,.
\end{align}
We introduce the operators that create a pair of mesons in a $C=+1$ state as in 
\begin{align}
 (\Bar{D}^0\bm{D}^0)_+={}&\Bar{D}^{0}\bm{D}^0+D^{0}\Bar{\bm{D}}^0 \,, \\
 (D^-\bm{D}^+)_+={}&D^{-}\bm{D}^+ + D^{+}\bm{D}^- \,,
\end{align}
which can also be recast in a doublet 
\begin{equation}
 (\Bar{D}\bm{D})_+=\begin{pmatrix}
 (\Bar{D}^0\bm{D}^0)_+\\
 (D^-\bm{D}^+)_+ 
 \end{pmatrix} \,.
 \label{eq:DD+}
\end{equation}
With these definitions the currents are
\begin{align}
\bm J_S={}&(\Bar{D}^0\bm{D}^0)_+-(D^-\bm{D}^+)_+ \,, \\
\bm J_T^3={}&-(\Bar{D}^0\bm{D}^0)_+ - (D^-\bm{D}^+)_+ \,.
\end{align}
The kinetic part of the non-relativistic Lagrangian is

\begin{align}
\label{Lkin}
 \mathcal{L}_{kin}&=D^\dagger\left(i\partial_t+\frac{\nabla^2}{2m_D}-\begin{pmatrix}
 \Delta_1 & 0\\
 0 & 0\\
 \end{pmatrix}\right)D+\Bar{D}^\dagger\left(i\partial_t+\frac{\nabla^2}{2m_D}-\begin{pmatrix}
 0 & 0\\
 0 & \Delta_1\\
 \end{pmatrix}\right)\Bar{D}+\notag\\ &+ \bm{D}^\dagger\left(i\partial_t+\frac{\nabla^2}{2m_{D^*}}-\begin{pmatrix}
 \Delta_2 & 0\\
 0 & 0\\
 \end{pmatrix}\right)\bm{D}+
 \Bar{\bm{D}}^\dagger\left(i\partial_t+\frac{\nabla^2}{2m_{D^*}}-\begin{pmatrix}
 0 & 0\\
 0 & \Delta_2\\
 \end{pmatrix}\right)\Bar{\bm{D}}+\notag\\&+\bm X^\dagger\left(i\partial_t+\frac{\nabla^2}{2(m_D+m_{D^*})}-m_F\right)\, \bm X \,.
\end{align}
In the parentheses we have the kinetic and mass terms, for which we have defined the zero at the mass threshold of the neutral $\Bar{D}^0\bm{D}^0$ pair. Thus, the mass parameters of the charged mesons are $\Delta_1=m_{D^+}-m_{D^0}\simeq 5 \text{ MeV}$ and $\Delta_2=m_{D^{*+}}-m_{D^{*0}}\simeq 3 \text{ MeV}$, and the one for the $X$ is $m_F=m_X-(m_{D^0}+m_{D^{*0}})$.

Considering the conservation of isospin in strong interactions, the interaction Lagrangian is given by
\begin{equation}
 \mathcal{L}_{\rm int}=-\frac{\lambda_S}{2}\bm J_S^{\dagger} \bm J_S-\frac{\lambda_T}{2}\bm J_T^{a,\dagger} \bm J_T^{a}-\frac{g}{\sqrt{2}}\bm X^\dagger\,\bm J_S+{\rm h.c.} \,.
 \label{eq:lag_int}
\end{equation}
The normalization factors are chosen for later convenience. The first two terms describe the mutual interactions between mesons, while the last term refers to the coupling between the isospin-singlet mesonic current and the elementary $X$.

We stress that here we limit ourselves to the instance where the elementary state is a pure isospin singlet. This is done for purely pedagogical reasons. The general expectation is to have both an isosinglet and an isotriplet. The complete effective theory should include both, with potentially two different couplings, $g$.

The interaction part involving neutral currents ($a=3$) can be made explicit and rewritten more conveniently using the vector $(\Bar{D}\bm{D})_+$ introduced in \eqref{eq:DD+}

\begin{align}
 \mathcal{L}_{\rm int}\supset{}&-\frac{\lambda_S}{2}(\Bar{D}\bm{D})_+^\dagger\begin{pmatrix}
 1 & -1\\
 -1 & 1
 \end{pmatrix}(\Bar{D}\bm{D})_+ -\frac{\lambda_T}{2}(\Bar{D}\bm{D})_+^\dagger\begin{pmatrix}
 1 & 1\\
 1 & 1
 \end{pmatrix}(\Bar{D}\bm{D})_+ \notag
 \\&-\frac{g}{\sqrt{2}}\bm X^\dagger(\Bar{D}^0\bm{D}^0)_+ + \frac{g}{\sqrt{2}} \bm X^\dagger(D^-\bm{D}^+)_+ \,.
 \label{eq:lag_int0}
\end{align}
The diagonal entries of the matrices correspond  the interactions where mesons do not change charge (i.e., $\Bar{D}^0\bm{D}^0 \to \Bar{D}^0\bm{D}^0$ or $D^-\bm{D}^+ \to D^-\bm{D}^+$), while the off-diagonal terms describe interactions between neutral and charged mesons. The last two terms describe the interactions between $X$ and the mesonic states (neutral and charged) with positive charge conjugation.

We need to describe the scattering between mesons in the $C=+1$ channel. Thus, we define the mesonic states with positive charge conjugation:
\begin{eqnarray}
 |1\rangle&=&\frac{1}{\sqrt{2}}\left(|\Bar{D}^0\bm{D}^0\rangle+|D^0\Bar{\bm{D}}^0\rangle\right)\,, \\
 |2\rangle&=&\frac{1}{\sqrt{2}}\left(|D^-\bm{D}^+\rangle+|D^+\bm{D}^-\rangle\right) \,.
\end{eqnarray}
These states are symmetric under charge conjugation and are interpolated by the operators $(\Bar{D}^0\bm{D}^0)_+$ and $(D^-\bm{D}^+)_+$.

In equation \eqref{eq:lag_int0}, we have already rewritten $\mathcal{L}_{\rm int}$ in terms of the operators we need, so we can directly derive the Feynman rules for the meson-meson interactions. We define the interaction matrix, $\Lambda$, as
\begin{equation}
 \Lambda=-i\begin{pmatrix}
 \lambda_T+\lambda_S && \lambda_T-\lambda_S \\
 \lambda_T-\lambda_S && \lambda_T+\lambda_S
 \end{pmatrix}=-i\left(\lambda_S\,\mathcal{P}_S+\lambda_T\,\mathcal{P}_T\right) \,,
\label{119}
\end{equation}
where $\mathcal{P}_i$ are the projectors onto the corresponding isospin states,
\begin{equation}
 \,\mathcal{P}_S=\begin{pmatrix}
 1 & -1 \\
 -1 & 1 \\
 \end{pmatrix}\,, \qquad \,\mathcal{P}_T=\begin{pmatrix}
 1 & 1 \\
 1 & 1 \\
 \end{pmatrix} \,.
\end{equation}
The matrix $\Lambda_{ij}$ provides the coupling between the initial state $|i\rangle$ and the final state $|j\rangle$. The factor $1/2$ in \eqref{eq:lag_int} is inserted so that the coupling is simply given by combinations of $\lambda_i$, with no extra numerical factors. Similarly, for the $X$, we can introduce the vector $G$ as
\begin{equation}
 G=-i\,g\,\begin{pmatrix}
 1\\
 -1
 \end{pmatrix} \,,
\end{equation}
such that the upper component corresponds to the coupling with neutral mesons, while the lower one refers to the coupling with charged mesons.

\subsection{ $D\bar D^*\to D\bar D^*$ scattering amplitude}
\begin{figure}
\centering
\includegraphics[width=0.99\linewidth]{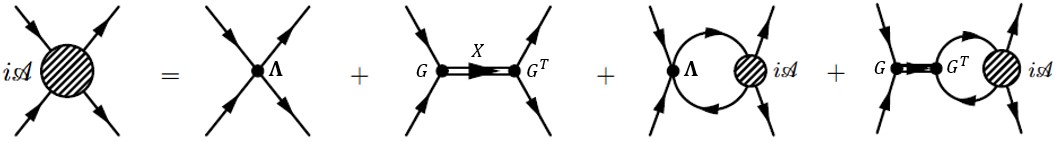}
\caption{Diagrammatic representation of the Schwinger-Dyson equation for the complete resummed amplitude. Single solid lines are $D$ or $D^*$ meson propagators, while double lines represent the $X$ propagator. In the loops, both neutral and charged meson pairs run.}
\label{fig:Amplitudes}
\end{figure}

In the non-relativistic theory, the scattering amplitude matrix can be computed non-perturbatively. The simplest way to perform this computation is to use a Schwinger-Dyson type equation, which can be proven diagrammatically. Denoting with a shaded circle the complete resummed amplitude for a given initial and final state, for example, for neutral-to-neutral scattering, this satisfies the equation reported in Fig.~\ref{fig:Amplitudes}. This equation can be written compactly as
\begin{equation}
 i\,\mathcal{A}={\Lambda_{\rm tree}}+{\Lambda_{\rm tree}}\,L(E)\,i\mathcal{A} \,,
 \label{eq:schwing-dyson}
\end{equation}
where $\mathcal{A}$ is the amplitude.
We have also introduced the tree-level scattering matrix ${\Lambda_{\rm tree}}$, which includes the first two terms of Fig.~\ref{fig:Amplitudes}
\begin{equation}
{\Lambda_{\rm tree}} = \Lambda+G\frac{i}{E-m_F+i\epsilon}G^T=-i\left[\left(\lambda_S+\frac{g^2}{E-{m_F}+i\epsilon}\right)\mathcal{P}_S+\lambda_T\,\mathcal{P}_T\right] \,.
\end{equation}
The $\Lambda$ term accounts for the short distance interactions between mesons, while the second term describes the interaction mediated by the $X$. All in all, the matrix $\big({\Lambda_{\rm tree}}\big)_{ij}$ describes the tree-level scattering between the initial state $|i\rangle$ and the final state $|j\rangle$, as induced by both the short distance contributions as well by the elementary $X$. Next, we introduce the loop integral $L(E)$, which depends on the product of the mesonic propagators
\begin{equation}
 \Pi_{\scriptscriptstyle D}(E,\bm{q})=\begin{pmatrix}
 \frac{i}{E-\frac{q^2}{2m_{D}}+i\epsilon} & 0\\
 0 & \frac{i}{E-\frac{q^2}{2m_D}-\Delta_1+i\epsilon}
 \end{pmatrix} \!, \;\,\,
 \Pi_{\scriptscriptstyle\boldsymbol{D}}(E,\bm q)=\begin{pmatrix}
 \frac{i}{E-\frac{q^2}{2m_{D^*}}+i\epsilon} & 0\\
 0 & \frac{i}{E-\frac{q^2}{2m_{D^*}}-\Delta_2+i\epsilon}
 \end{pmatrix} \!,
\end{equation}
\begin{equation}
 L(E)=\int\!\frac{d^4 q}{(2\pi)^4}\Pi_{\scriptscriptstyle \bm{D}}\!\left(E-q_0,-\bm q\right)\Pi_{\scriptscriptstyle D}\!\left(q_0,\bm q\right)=i\frac{m}{2\pi}\begin{pmatrix}
 \sqrt{-2m E-i\epsilon} & 0\\
 0 & \sqrt{-2m(E-\Delta)-i\epsilon}\\
 \end{pmatrix} \,,
 \label{eq:loop}
\end{equation}
where $m$ is the reduced mass of the $D^0\Bar{{D}}^{*0}$ system, which we assume is the same as that of the $D^-{D}^{*+}$ system, and $\Delta=\Delta_1+\Delta_2\simeq 8\text{ MeV}$ (a flavor symmetry breaking parameter). 

Since we are considering only neutral states, the off-diagonal terms (which correspond to ``charged" loops) are null. The integral is linearly divergent and must be regularized. Here, we report only the finite part. For convenience, we also define the complex momenta 
\begin{equation}
 \kappa_0(E)=\sqrt{-2m E-i\epsilon}\,,\quad \kappa_+(E)=\sqrt{-2m(E-\Delta)-i\epsilon} \,.
\end{equation}
The solution to equation \eqref{eq:schwing-dyson} is therefore 
\begin{equation}
 i\mathcal{A}=\left(1-{\Lambda_{\rm tree}}\,L(E)\right)^{-1}{\Lambda_{\rm tree}}=\left({\Lambda_{\rm tree}^{-1}}-L(E)\right)^{-1} \,,
\end{equation}
which explicitly gives 
\begin{equation}
 \mathcal{A}(E)=\left\{-\left[\left(\lambda_S+\frac{g^2}{E-{m_F}+i\epsilon}\right)\mathcal{P}_S+\lambda_T\,\mathcal{P}_T\right]^{-1}+\frac{m}{2\pi}\begin{pmatrix}
 \kappa_0(E) & 0\\
 0 & \kappa_+(E)\\
 \end{pmatrix}\right\}^{-1} \,.
\end{equation}
The non-relativistic scattering amplitude is related to \(\mathcal{A}\) by the relation
\begin{equation}
 \mathcal{A}(E)=\frac{2\pi}{m}f(E) \,,
\end{equation}
so that
\begin{equation}
 f(E)=\left\{-\frac{2\pi}{m}\left[\left(\lambda_S+\frac{g^2}{E-{m_F}+i\epsilon}\right)\mathcal{P}_S+\lambda_T\,\mathcal{P}_T\right]^{-1}+\begin{pmatrix}
 \kappa_0(E) & 0\\
 0 & \kappa_+(E)\\
 \end{pmatrix}\right\}^{-1} \,.
\end{equation}
For simplicity, we can redefine (with an abuse of notation)\footnote{In what follows, $[\lambda_i]=E^{-1}$ and $[g]=E^0$.}
\begin{equation}
    \lambda_i\mapsto\frac{2\pi}{m}\lambda_i\,,\quad g\mapsto\sqrt{\frac{2\pi}{m}}g
    \label{eq:ridef}
\end{equation}
to eliminate the factor $2\pi/m$ from the first term as well
\begin{equation}
 f(E)=\left\{-\left[\left(\lambda_S+\frac{g^2}{E-{m_F}+i\epsilon}\right)\mathcal{P}_S+\lambda_T\,\mathcal{P}_T\right]^{-1}+\begin{pmatrix}
 \kappa_0(E) & 0\\
 0 & \kappa_+(E)\\
 \end{pmatrix}\right\}^{-1} \,.
 \label{1130}
\end{equation}
This is the most general expression for the scattering amplitude that accounts for short-range mesonic scattering, including the charged contribution, and allows for the existence of an elementary resonance. 

In the following sections, we will analyze two limiting cases: the absence of the elementary $X$ ($g=0$) or the absence of short-range interactions ($\Lambda=\bm{0}$).

\subsubsection{Purely molecular model $(g=0)$}
\label{sec:pure_mol}
The scattering amplitude in absence of the elementary resonance (i.e., $g=0$) reduces to
\begin{equation}
 f(E)=\left[-\left(\lambda_S\mathcal{P}_S+\lambda_T\,\mathcal{P}_T\right)^{-1}+\begin{pmatrix}
 \kappa_0(E) & 0\\
 0 & \kappa_+(E)\\
 \end{pmatrix}\right]^{-1} \,.
 \label{1131}
\end{equation}
More explicitly
\begin{equation}
 f(E)=\frac{1}{D_0(E)}\begin{pmatrix}
 \lambda_S(4\,\kappa_+(E)\,\lambda_T-1)-\lambda_T & \lambda_S-\lambda_T\\
 \lambda_S-\lambda_T& \lambda_S(4\,\kappa_0(E)\,\lambda_T-1)-\lambda_T\\
 \end{pmatrix} \,,
 \label{f_E_molecule}
\end{equation}
where 
\begin{equation}
 D_0(E)=1-(\lambda_T+\lambda_S)(\kappa_0+\kappa_+)+4\kappa_0\kappa_+\lambda_S\lambda_T \,,
\end{equation}
with $\kappa_0(E)=\sqrt{-2mE-i\epsilon}$ and $\kappa_+(E)=\sqrt{-2m(E-\Delta)-i\epsilon}$.

The interaction parameters $\lambda_S$ and $\lambda_T$ could be such that there is a pole in the scattering amplitude just below the $\Bar{D}^0 D^{*0}$ threshold, which corresponds to the physical $X$. The condition $D_0(-B)=0$, where $B>0$ is the binding energy of the bound state, leads to the linear equation (to further simplify the problem, let's approximate $\kappa_+(-B)\simeq \kappa_+(0)=\sqrt{2m\Delta}$)
\begin{equation}
 \sqrt{2mB}=\frac{1-\sqrt{2m\Delta}\,(\lambda_T+\lambda_S)}{\lambda_S + \lambda_T - 
 4\sqrt{2m\Delta}\lambda_S\lambda_T} \,. \label{eq:gamma}
\end{equation}

As far as the scattering length and effective range are concerned, as long as the energy is lower than that of the charged threshold, $E<\Delta$, the only possible scattering is that of neutral mesons, whose corresponding amplitude is
\begin{equation}
    f_{11}(E)=\frac{\lambda_S(4\,\kappa_+(E)\,\lambda_T-1)-\lambda_T}{1-(\lambda_T+\lambda_S)(\kappa_0+\kappa_+)+4\kappa_0\kappa_+\lambda_S\lambda_T} \,.
    \label{eq:f_11}
\end{equation}
Assuming low-energy scattering near the neutral threshold, we can expand the inverse of the amplitude $f_{11}(E)$ in powers of the relative momentum $k$ between the mesons, obtaining
\begin{equation}
 f^{-1}_{11}\left(E=\frac{k^2}{2m}\right)=-\frac{1}{a}-ik+\frac{1}{2}r_0\,k^2+O\left(k^4\right) \,,
\end{equation}
where the following expressions for scattering length and effective range are found\footnote{Compare with (2.21) in \cite{Cohen}, identifying
\begin{equation}
 a_{11}=a_{22}=\frac{4\lambda_S\lambda_T}{\lambda_S+\lambda_T}\,,\quad\quad a_{12}=\frac{4\lambda_S\lambda_T}{\lambda_S-\lambda_T} \,. \notag
\end{equation}} 
\begin{equation}
 a^M=\frac{\lambda_T-\lambda_S(4\lambda_T\sqrt{2m\Delta}-1)}{1-(\lambda_S+\lambda_T)\sqrt{2m\Delta}}\,, \qquad\quad r_0^M=-\frac{1}{\sqrt{2m\Delta}}\left(\frac{\lambda_S-\lambda_T}{\lambda_S+\lambda_T-4\lambda_S\lambda_T\sqrt{2m\Delta}}\right)^2 \,.
 \label{eq:as_r0}
\end{equation}
Despite the fact that in this purely molecular scenario only contact interactions are included, we obtain a negative scattering range --- contrary to the single threshold analysis performed by Weinberg. This negative contribution to $r_0$ is indeed related to the inclusion of the charged threshold and it vanishes in the limit $\Delta\to+\infty$, see discussion in Section~\ref{sez:r0charge}. This is to be expected, as when the charged mesons are substantially heavier than the neutron ones, they could be further integrated out of the effective theory, ending up into higher derivative corrections to the contact interaction of the neutral mesons. These induce nothing but the corrections to Weinberg's formula reported in Eq.~\eqref{r0_Weinberg}, modulo that the cutoff is now $\Delta$. Moreover, recall that pion interactions also push $r_0$ in the negative range (Section~\ref{correpion}).
The absence of any additional threshold with respect to $np$ in the deuteron case keeps $r_0$ positive, as described in Section~\ref{r0sign}.

Interestingly if $\lambda_S=\lambda_T$, when the matrix $\Lambda$ in~\eqref{119} becomes diagonal, we are left with $r_0=0$, plus corrections, as a result of~\eqref{1131} being diagonal. 

The scattering amplitude \eqref{f_E_molecule} is markedly different from the Flatté parametrization used by LHCb \cite{LHCb:2020xds}. The denominator $D_0(E)$ does not have the same energy dependence as in \eqref{ikappa}, and there is an energy dependence in the numerators of $f_{11}(E)$ and $f_{22}(E)$.\footnote{The difference between the two numerators would be evident in a $DD^* \to DD^*$ scattering process. However, the $X(3872)$ is tagged by the invariant mass of $J/\psi \pi^+\pi^-$ in inclusive $b$-hadron decays \cite{LHCb:2020xds}. This introduces an unknown smooth energy dependence in the numerator even in the Flatté case.} 

In order to test the molecular hypothesis, the lineshape of the $X$ must be fitted using \eqref{f_E_molecule}. However, as mentioned in the previous section, the $r_0^M$ obtained from the fit is subject to the effect of the charged threshold, while Weinberg's analysis applies to single-channel scattering only. The prescription provided by \cite{r0_isospin} to handle this problem is that of subtracting the isospin symmetry-violating terms related to the charged channel from $a^M$ and $r_0^M$ before calculating $Z$. If we apply this prescription to $r_0$, we obtain a trivial result, since 
\begin{equation}
 \Delta r^I_M=-r_0^M=\frac{1}{\sqrt{2m\Delta}}\left(\frac{\lambda_S-\lambda_T}{\lambda_S+\lambda_T-4\lambda_S\lambda_T\sqrt{2m\Delta}}\right)^2 \,,
\end{equation}
thus 
\be 
r_0^\prime=r_0^M+\Delta r^I_M=0 \,.
\ee 
This is not surprising, as in the purely molecular hypothesis ($Z=0$), we expect from Weinberg's formula \eqref{err0} that $r_0=0$, plus higher-order corrections. Therefore, extracting $Z$ from the parametrization in Eq.~\eqref{f_E_molecule} is meaningless, as it already assumes the molecular nature of the $X$. One should rather consider the goodness of the fit.

In principle, to apply Weinberg's criterion in the molecular case, $r_0$ should be fitted directly from experimental data using the universal, model-independent, expression of the low-energy scattering amplitude \eqref{err0bethe}
\begin{equation}
f=\frac{1}{-1/a+\frac{1}{2}r_0k^2-ik} \,,
\label{eq:universal}
\end{equation}
as was done for $np$ scattering \cite{Bethe, np_r0}. We will refer to the $r_0$ obtained this way as the universal $r_0^U$. From $r_0^U$ the isospin effect has to be removed, following~\cite{r0_isospin}
\begin{equation}
 r_0^\prime=r_0^U+\Delta r^I_M=r_0^U+\frac{1}{\sqrt{2m\Delta}}\left(\frac{\lambda_S-\lambda_T}{\lambda_S+\lambda_T-4\lambda_S\lambda_T\sqrt{2m\Delta}}\right)^2 \,, \label{eq:r0prime}
\end{equation}
to exclude the effects of the coupled-channel. If it turns out that $r_0^\prime<0$ (driven by a negative $r_0^U$ value), it has to be compared with \eqref{eq:r_o_esposito} to see if pion corrections could have led to such a result. Conversely if $r_0^\prime>0$, it could be a signal of the molecular nature of the $X$. We want to emphasize that the scattering amplitude can be approximated by Eq.~\eqref{eq:universal} only near the neutral threshold, where higher-order terms in the expansion of $k\cot\delta$ can be neglected (see Sec.~\ref{ar}). In contrast, Eq.~\eqref{eq:f_11} remains valid wherever the effective theory applies.

Recently, an extension of the molecular model proposed here that also includes corrections due to the pion has been studied in \cite{Dong_conf}. Using this upgrade to fit the lineshape of the $X$ instead of the Flatté amplitude, already discussed in Section \ref{zfls}, leads to an $r_0>0$. However, what is presented seems to contain an inaccuracy in how to treat the isospin breaking effect. This effect should be taken into account by adding to the $r_0$ extrapolated by data, $\Delta r^I_M$
and not the one predicted for the Flatté case \cite{r0_isospin}
\begin{equation}
 \Delta r^I_F=\sqrt{\frac{m_+}{2m^2\Delta}}
\end{equation}
where $m_+$ is the reduced mass of the charged system (which we have assumed to be equal to $m$ for simplicity).

\subsubsection{Purely compact model, or Flatté limit $(\Lambda={0})$}
\label{purelycompact}
The case $\Lambda=\bm{0}$ does not correspond to reality because it is known that the $D$ and $D^*$ mesons can interact through short-range interactions. However, what will be described here can be regarded as the limiting case in which the coupling to the resonance dominates over other types of interaction, at least in the energy range in which the effective theory is valid, and therefore the scattering proceeds almost exclusively through the channel of the $X$. In this case, the scattering amplitude reduces to
\begin{equation}
 f(E)=\left[-\left(\frac{g^2}{E-{m_F}+i\epsilon}\,\mathcal{P}_S\right)^{-1}+\begin{pmatrix}
 \kappa_0(E) & 0\\
 0 & \kappa_+(E)\\
 \end{pmatrix}\right]^{-1} \,.
 \label{eq:f(E)_flatte}
\end{equation}
We can use the same results from the previous section with the substitution
\begin{eqnarray}
 \lambda_S\to\frac{g^2}{E-{m_F}+i\epsilon} \,, \qquad \lambda_T\to0 \,,
\end{eqnarray}
because the $X$ couples only to the isosinglet channel. If we consider the component $f_{11}(E)$, the amplitude reduces to
\begin{equation}
 f_{11}(E)=-\frac{g^2}{E-{m_F}-g^2\left[\kappa_0(E)+\kappa_+(E)\right]} \,,
 \label{flattult}
\end{equation}
which is indeed the Flatté amplitude \cite{Flatte}, as in~\eqref{tolhcb}, and where $\kappa_0(E)=\sqrt{-2mE-i\epsilon}$ and $\kappa_+(E)=\sqrt{-2m(E-\Delta)-i\epsilon}$.

As in the molecular case, we consider the range $E<\Delta$ and expand $f^{-1}_{11}(E)$ in powers of the relative momentum $k$, obtaining
\begin{equation}
 \frac{1}{a^F}=-\frac{m_F}{g^2}-\sqrt{2 m \Delta}\qquad\quad r_0^F=-\frac{1}{m\,g^2}-\sqrt{\frac{1}{2 m \Delta}}
 \label{eq:as_r0_flatte}
\end{equation}
The effective range in the Flatté amplitude is forced to be negative. However, this fact is not connected to the limit $Z \to 1$ of \eqref{err0}, since the coupling $g$ that appears in the Flatté amplitude is neither a function of $B$ nor of $Z$. The expression of $r_0^F$ derived here gives a negative and finite result, unlike Weinberg's expression \eqref{err0}, which becomes infinite in the case of $Z \to 1$.

We have also rederived the term due to the coupled-channel effects~\cite{r0_isospin} (approximating the reduced mass of the charged channel $m_+$ as that of the neutral one), i.e.,
\begin{equation}
 \Delta r^I_F=\sqrt{\frac{1}{2m\Delta}} \,,
\end{equation}
which contributes to the value of $r_0$ and depends on the isospin breaking parameter $\Delta$. According to \cite{r0_isospin}, this value should be added to the experimental value of $r_0$ before extracting $Z$ to account for the fact that Weinberg's analysis is not meant for coupled channels. However, as mentioned several times, this prescription does not make sense for the Flatté case since, in this model, there are no bound states. Again, it is much more solid to simply try and fit the data, and then assess the goodness of fit.

\begin{figure}[t]
 \centering
 \includegraphics[width=0.5\linewidth]{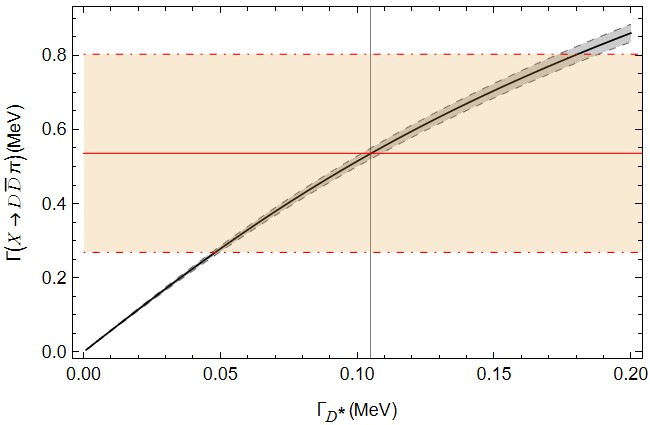}
 \caption{The black solid line represents the central value of $\Gamma(X\to D\Bar{D}\pi)$ calculated using \eqref{deca} and $g^2_X=g_c^2$ from \eqref{eq:gc} as a function of $\Gamma_{D^*}$; the red solid line is the measured value from PDG \cite{ParticleDataGroup:2024cfk}. The bands show the one-sigma intervals.}
 \label{fig:GammavsGamma}
\end{figure}
The expressions \eqref{eq:as_r0_flatte} for $a^F$ and $r_0^F$ coincide with those used by LHCb \eqref{eq:a0_LHCb}, under the assumption $m_+=m$ and identifying $g_{\text{LHCb}}=2 g^2$. As discussed above, we have that the coupling $g$ in formula~\eqref{1130} corresponds to the coupling in the Lagrangian (indicated with the same letter) multiplied by $\sqrt{m/2\pi}$ thus
\be
\left(\sqrt{\frac{m}{2\pi}} g \right)^2=\frac{g_{\rm LHCb}}{2} \,,
\label{redef}
\ee
comparing~\eqref{finlhc} with~\eqref{flattult} which gives 
\be
g^2=\frac{\pi}{m} g_{\rm LHCb} \,.
\ee
This coupling here is what we would have called the $g_X$ coupling in Section~\ref{laonly}. However, as commented above, the Flatt\`e is describing the purely compact state coupled to the continuum, therefore this analysis finds that 
\be
g_c^2=\frac{\pi}{m} g_{\rm LHCb} \,,
\ee
where $g_c$ was introduced in Eq.~\eqref{418} as
\be
\langle \alpha|V|\mathfrak X\rangle\equiv g_c \,, \qquad g_X \simeq g_c \,.
\ee
If we go back to Section~\ref{laonly}, and we decide to study the $\Gamma(X\to D\bar D\pi)$ decay assuming that the $X$ is a purely elementary state, then we have to use this coupling $g_c$, which does not depend on $B$. Using the LHCb measurement reported in Eq.~\eqref{gLHCbfit}, one estimates $g_c$ to be
\begin{equation}
 g_c = (1.87 \pm 0.03)\times 10^{-2}\,\text{MeV}^{-1/2} \,.
 \label{eq:gc}
\end{equation}
We obtain therefore the result for $\Gamma(X\to D\bar D\pi)$ as shown in Fig.~\ref{fig:GammavsGamma}. 

We can provide an estimate of $\Gamma_{D^{*0}}$ using $g_c^2$ for the calculation of the width $\Gamma(X \to D\Bar{D}\pi)$ in \eqref{deca}, whose PDG value is given by $\Gamma(X \to D\Bar{D}\pi) = (54 \pm 27) \text{ keV}$ \cite{ParticleDataGroup:2024cfk}. This corresponds to 
\begin{equation}
 \Gamma_{D^{*0}} = 105 \pm 70 \text{ keV} \,,
\end{equation} 
a value which is compatible with calculations reported in several articles~\cite{Gamma2011, Gamma2021, Gamma2024}, and with the measured value for the charged counterpart $\Gamma_{D^{*\pm}}=(83.4\pm1.8) \text{ keV}$~\cite{ParticleDataGroup:2024cfk}.

Summarizing, using the LHCb data on the lineshape of the $X$, an excellent agreement with the $X\to D\bar D\pi$ decay rate is found under the hypothesis that $X$ is purely elementary (i.e. $g_X=g_c$). This happens in correspondence to a value of $\Gamma_{D^{*0}}$ which is very close to the one found experimentally for $\Gamma_{D^{*\pm}}$ (we remind that for the former the PDG provides only an upper bound at 2.1~MeV). 

\subsubsection{From Flatté to the molecular amplitude}
The scattering amplitudes discussed in the previous sections exhibit two different behaviors depending on the nature of the $X(3872)$. However, starting from the Flatté amplitude \eqref{flattult}, one can recover the molecular one, Eq.~\eqref{eq:f_11}, by taking the limit $g^2 \to \infty$. Since the $X$ mediates only the isoscalar interaction, the result of the limit should be compared with Eq. \eqref{eq:f_11} setting $\lambda_T = 0$.

Consider the interaction terms
\begin{equation}
    \mathcal{L}_{\text{int}} = -\frac{g}{\sqrt{2}} \bm{X}^\dagger (\bar{D}^0\bm{D}^0)_+ + \frac{g}{\sqrt{2}} \bm{X}^\dagger (\bar{D}^-\bm{D}^+)_+ + \text{h.c.} \,,
    \label{eq:lag_limit}
\end{equation}
and redefine
\begin{equation}
    \bm{X}^\prime = g\,\bm{X} \,.
\end{equation}
Consequently, the kinetic terms for the field $\bm{X}^\prime$ becomes
\begin{equation}
    \mathcal{L}_{\text{kin}} \supset \bm{X}^{\prime\,\dagger} \frac{1}{g^2} \left(i\partial_t + \frac{\nabla^2}{2(m_D+m_{D^*})} - m_F \right) \bm{X}^\prime \,.
    \label{eq:lag_kin_limit}
\end{equation}
Taking the strong coupling limit, $g^2\to\infty$, but keeping $m_F/g^2$ finite, the field becomes non-dynamical and its equations of motion reduce to
\begin{equation}
    \frac{m_F}{g^2} \bm{X}^\prime + \frac{1}{\sqrt{2}}(\bar{D}^0\bm{D}^0)_+ - \frac{1}{\sqrt{2}}(D^-\bm{D}^+)_+ = 0 \,.
\end{equation}
The parameter $m_F$ alone has no physical meaning since the physical mass depends on the ratio $m_F/g^2$ (see Eq. \eqref{eq:as_r0_flatte}). Solving for $\bm{X}^\prime$ and substituting back into Eqs. \eqref{eq:lag_limit} and \eqref{eq:lag_kin_limit}, we obtain
\begin{equation}
    \mathcal{L}_{\text{int}} = \frac{g^2}{2m_F} (\bar{D}\bm{D})^\dagger_+ 
    \begin{pmatrix}
        1 & -1 \\
        -1 & 1
    \end{pmatrix}
    (\bar{D}\bm{D})_+ \,.
\end{equation}
This is equivalent to the Lagrangian \eqref{eq:lag_int0} without the $\bm{X}$ field, by setting
\begin{equation}
    \lambda_S = -\frac{g^2}{m_F} \,,
\end{equation}
and $\lambda_T = 0$ (as we are considering the special case where the $X$ only mediates scattering in the isosinglet channel). The resulting scattering amplitude for neutral mesons is\footnote{In this formula and the following one, the parameters $\lambda_S$ and $g$ are redefined as in Eq. \eqref{eq:ridef}.}
\begin{equation}
    f_{11}(E) = \frac{1}{-\lambda_S^{-1} + \sqrt{-2mE - i\epsilon} + \sqrt{-2m(E-\Delta) - i\epsilon}} \,,
\end{equation}
which is equivalent to Eq. \eqref{eq:f_11} with $\lambda_T=0$. The same result could have been obtained starting from the Flatt\'e amplitude
\begin{equation}
    f_{11}(E) = -\frac{g^2}{E - m_F - g^2 [\kappa_0(E) + \kappa_+(E)]} \,,
\end{equation}
and taking the limit $g^2\to\infty$ with $m_F/g^2$ finite.

We want to emphasize two important points. 
%After taking the limit, the only contribution to $r_0$ comes from the presence of the charged threshold. 
First of all, it is experimentally meaningless to measure an infinite value for $g^2$, making it unclear at what point one can definitively claim that the system no longer corresponds to a compact state but rather to a molecular one.

Moreover, after taking the limit described above, the only contribution to $r_0$ comes from the presence of the charged threshold. From an analysis of the LHCb data~\cite{LHCb:2020xds}, one finds $r_0 \neq 0$, even after subtracting this contribution. This points to the fact that indeed the coupling $g^2$ is finite.

It is also possible to start from a Lagrangian with only quartic interactions and obtain an equivalent description by introducing a \textit{non-dynamical} field with only trilinear interactions \cite{Braaten:2020nmc}. Consider the interaction Lagrangian
\begin{equation}
    \mathcal{L}_{\rm int}=-\frac{\lambda_S}{2} (\bar{D}\bm{D})^\dagger_+ 
    \begin{pmatrix}
        1 & -1 \\
        -1 & 1
    \end{pmatrix}
    (\bar{D}\bm{D})_+\,,
    \label{eq:l_int_mol}
\end{equation}
which is equivalent to the one discussed in Sec. \ref{sec:pure_mol}, with $\lambda_T=0$. We can define a new field $\bm{X}^\prime$ as
\begin{equation}
    \bm{X}^\prime=\frac{\lambda_S}{\sqrt{2}}\bigg((\bar{D}^0\bm{D}^0)_+-(D^-\bm{D}^+)_+\bigg) \,,
\end{equation}
and it is straightforward to see that the interaction Lagrangian \eqref{eq:l_int_mol} reduces to 
\begin{equation}
    \mathcal{L}_{\rm int}=\frac{1}{\lambda_S}\bm{X}^{\prime\,\dagger}\bm{X}^\prime-\frac{1}{\sqrt{2}}(\bar{D}^0\bm{D}^0)_+^\dagger\bm{X}^\prime+
    \frac{1}{\sqrt{2}}(D^-\bm{D}^+)_+\bm{X}^\prime+\text{h.c.}\,.
    \label{eq:l_int_molX}
\end{equation}

A Flatté-like amplitude can also be used to describe the molecular scenario without requiring any coupling to be sent to infinity. As we did for the compact hypothesis, we can introduce a \textit{dynamical} field for the $X(3872)$ even in the molecular picture, provided the necessary precautions are taken \cite{Kaplan:1996nv,Kaplan:1999qa}. Furthermore, this approach allows the inclusion of short-distance contributions to $r_0$, which are essential for determining the nature of the particle.

Suppose we aim to describe a molecular \textit{isosinglet} $X(3872)$. In this case, we can introduce a field $\bm{X}_m$ that couples to the mesonic current through a conventional trilinear interaction
\begin{equation} 
\mathcal{L}_{int}= -\frac{g}{2}\bm{X}_m^\dagger(\bar{D}^0\bm{D}^0)_++\frac{g}{2}\bm{X}_m^\dagger(D^-\bm{D}^+)_++\text{h.c.}\,,
\end{equation} 
but its kinetic term needs the overall minus sign\cite{Kaplan:1996nv}
\begin{equation} \mathcal{L}_{kin,\,X}=-\bm{X}_m^\dagger\left(i\partial_t+\frac{\nabla^2}{2(m_D+m_{D^*})}-m_F\right)\bm{X}_m \,. 
\end{equation}
The ‘wrong' sign is introduced to ensure a positive contribution to $r_0$ due to the (attractive) short-range interactions that bind the $X(3872)$. This fact can be easily proved considering that the minus sign changes the propagator's denominator, leading to the following expression for the scattering amplitude
\begin{equation}
    f_{11}(E) = \left[ -\left(\frac{g^2}{-(E - m_F + i\epsilon)}\mathcal{P}_S\right)^{-1} 
    + \begin{pmatrix}
        \kappa_0(E) & 0 \\
        0 & \kappa_+(E)
    \end{pmatrix} \right]^{-1}\,.
\end{equation}
This is equivalent to Eq.~\eqref{eq:f(E)_flatte} upon sending $g^2 \mapsto -g^2$. This implies that the scattering length and the effective range are given by
\begin{equation}
    \frac{1}{a} = \frac{m_F}{g^2} - \sqrt{2m\Delta}, \qquad\quad 
    r_0 = \frac{1}{mg^2} - \sqrt{\frac{1}{2m\Delta}}.
\end{equation}
Moreover, to ensure the positivity of the scattering length (i.e., that the pole corresponds to positive imaginary momentum), $m_F$ must be positive. In contrast, in Eq.~\eqref{eq:as_r0_flatte}, $m_F$ has to be negative. 
 
This can be seen as a practical application of Weinberg's criterion. The sign of $r_0$ highlights the difference between a compact state (with a positive kinetic term) and a bound state, which enters the Lagrangian as an auxiliary field (with an inverted kinetic term). 

We want to emphasize that the field $\bm{X}_m$ is introduced only to replace the \textit{short-range} interactions between mesons. This means that its contribution to the total $r_0$ must mimic the effect of short-distance physics which do not include pion exchange. The effects of the pion can be included perturbatively \cite{Braaten:2020nmc,Esposito:2023mxw}. 

\subsubsection{Summary on the scattering amplitudes}
\label{sommariols}

{\bf {\emph{Flatté amplitude. }}}Let us start with the Flatté amplitude. This amplitude can be derived from an effective theory that does not include quartic interactions between mesons but the mediation of an elementary $X$ with coupling constant $g$. If we define $m_F$ as the mass parameter of the elementary $X$ with respect to the neutral threshold (see \eqref{Lkin}), and $m$ as the reduced mass of the $\Bar{D}D^{*0}$ system (which we assume to be equal to that of $D^-D^{*+}$), the scattering amplitude takes the form 
\begin{equation}
 f(E)=-\frac{g^2}{E-{m_F}-g^2\left[\sqrt{-2mE-i\epsilon}+\sqrt{-2m(E-\Delta)-i\epsilon}\right]} \,.
\end{equation}
The expansion in powers of the relative momentum $k=\sqrt{2mE}$ provides
\begin{equation}
 \frac{1}{a^F}=-\frac{m_F}{g^2}-\sqrt{2 m \Delta}\,,\qquad\quad r_0^{F}=-\frac{1}{m\,g^2}-\sqrt{\frac{1}{2 m \Delta}} \,.
\end{equation}
The coupling $g$ between the resonance and the mesons is a purely phenomenological parameter and is not related in any way to $Z$ or the `binding energy' $B$ of a hypothetical bound state that is, by definition, not present in the theory. Indeed, $r_0^{{F}}$ is a finite quantity, whereas in \eqref{err0} $r_0 \to -\infty$ for $Z \to 1$. The coupling $g$ should be related to $g_c$ introduced in Eq.~\eqref{418} as the coupling between the compact $X$ and the continuum $DD^*$
\begin{equation}
g_c\equiv\langle \alpha|V|\mathfrak X\rangle \,,
\end{equation}
as done in Section \ref{purelycompact}. Consequently, the estimate of $Z$ obtained by considering the parameters of a Flatté amplitude does not appear to be meaningful.

{\bf {\emph{Molecular case. }}}To test the purely molecular hypothesis, the scattering amplitude that should be used is given by equation \eqref{f_E_molecule}
\begin{equation}
 f(E)=\frac{1}{D_0(E)}\begin{pmatrix}
 \lambda_S(4\,\kappa_+(E)\,\lambda_T-1)-\lambda_T & \lambda_S-\lambda_T\\
 \lambda_S-\lambda_T& \lambda_S(4\,\kappa_0(E)\,\lambda_T-1)-\lambda_T\\
 \end{pmatrix} \,,
\end{equation}
where 
\begin{equation}
 D_0(E)=1-(\lambda_T+\lambda_S)(\kappa_0+\kappa_+)+4\kappa_0\kappa_+\lambda_S\lambda_T \,,
\end{equation}
with $\kappa_0(E)=\sqrt{-2mE-i\epsilon}$ and $\kappa_+(E)=\sqrt{-2m(E-\Delta)-i\epsilon}$. In this expression, $\lambda_S$ and $\lambda_T$ are the couplings of the quartic interactions between mesons in the singlet and triplet channels (see Eq.~\eqref{eq:lag_int}), which must be fitted using the lineshape of the $X$. Once the interaction parameters are known, the binding energy of the bound state $B$ can be derived from the position of the zero of $D_0(E)$, using the formula
\begin{equation}
 \sqrt{2mB}=\frac{1-\sqrt{2m\Delta}\,(\lambda_T+\lambda_S)}{\lambda_S + \lambda_T - 
 4\sqrt{2m\Delta}\lambda_S\lambda_T} \,.
\end{equation}
Expanding around the neutral threshold, the following expressions for scattering length and effective range are found \eqref{eq:as_r0} 
\begin{equation}
 a^M=\frac{\lambda_T-\lambda_S(4\lambda_T\sqrt{2m\Delta}-1)}{1-(\lambda_S+\lambda_T)\sqrt{2m\Delta}} \,, \qquad\quad r_0^{M}=-\frac{1}{\sqrt{2m\Delta}}\left(\frac{\lambda_S-\lambda_T}{\lambda_S+\lambda_T-4\lambda_S\lambda_T\sqrt{2m\Delta}}\right)^2 \,.
\end{equation}

The value $r_0^M$ cannot be directly compared with Weinberg's formula \eqref{err0} since Weinberg's treatment does not include the effect of coupled channels (charged mesons), which is instead contained in $r_0^M$. According to \cite{r0_isospin}, to take this effect into account, one should add to an $r_0$ obtained directly from the lineshape all the terms $\Delta r^I$ dependent on the isospin breaking parameter $\Delta$ and compare this result with the Weinberg formula for $r_0$ \eqref{err0}. However, this would lead to a trivial result since in the molecular case, $\Delta r_M^I=-r_0^M$, thus $r_0^\prime=r_0+\Delta r_M^I=0$. This is what we expect from $Z=0$ unless there are higher-order corrections. To include these corrections, $r_0$ should be fitted directly from experimental data, using the universal, model-independent, expression of the low-energy scattering amplitude \eqref{err0bethe}
\begin{equation}
f=\frac{1}{-1/a+\frac{1}{2}r_0k^2-ik} \,,
\end{equation}
as was done for $np$ scattering \cite{Bethe, np_r0}. We will call the $r_0$ obtained in this way the universal $r_0^U$ and then calculate
\begin{equation}
 r_0^\prime=r_0^U+\Delta r^I_M=r_0^U+\frac{1}{\sqrt{2m\Delta}}\left(\frac{\lambda_S-\lambda_T}{\lambda_S+\lambda_T-4\lambda_S\lambda_T\sqrt{2m\Delta}}\right)^2 \,,
\end{equation}
to eliminate the effects of the coupled channel. If $r_0^\prime<0$, this value has to be compared with \eqref{eq:r_o_esposito} to see if pion corrections could have led to such a result or if another mechanism is at play. Conversely if $r_0^\prime>0$, it could be a signal of the molecular nature of the $X$, provided that there is a model that explains the obtained value \cite{Dong_conf}. Thus, the sign (and value) of $r_0^\prime$ can be used to determine the nature of the $X$, rather than $Z$.

\subsection{Correction to $r_0$ due to the charged threshold in molecular model}
\label{sez:r0charge}
The inclusion of the charged counterpart in the scattering of $\Bar{D}^0D^{*0}$ mesons has led to the appearance of an $r_0 \propto -1/\sqrt{2m\Delta} \simeq -1.6\,\text{fm}$, despite only contact interactions were included. We also note that the effect disappears in the limit $\Delta \to +\infty$, emphasizing the fact that the origin of this $r_0$ is due to the charged threshold. We provide an explanation of this in low energy scattering theory.

Analyzing the neutral open charm mesons channel in non-relativistic quantum mechanics, annihilation/creation of charged pairs is not allowed. A generic initial state $|\Psi\rangle$ can be written as
\begin{equation}
 |\Psi\rangle=|\Psi^0\rangle+|\Psi^+\rangle \,,
\end{equation}
where $|\Psi^0\rangle$ ($|\Psi^+\rangle$) is a generic state of neutral (charged) mesons. We also define the projectors $\mathcal{P}^0$ and $\mathcal{P}^+$ such that
\begin{equation}
 |\Psi^0\rangle=\mathcal{P}^0|\Psi\rangle \,, \qquad |\Psi^+\rangle=\mathcal{P}^+|\Psi\rangle \,.
\end{equation}
We denote $H$ as the general Hamiltonian which includes all the kinetic terms and interactions. Using the projectors, we can construct the Hamiltonians
\begin{eqnarray}
 H^0&=&\mathcal{P}^0\,H\,\mathcal{P}^0 \,, \\
 H^+&=&\mathcal{P}^+\,H\,\mathcal{P}^+ \,, \\
 H^{0+}&=&\mathcal{P}^0\,H\,\mathcal{P}^+ \,, \\
 H^{+0}&=&\mathcal{P}^+\,H\,\mathcal{P}^0 \,,
\end{eqnarray}
$H^0$ and $H^+$ contain the kinetic terms and the interactions within the neutral and charged subspaces, while $H^{0+}$ and $H^{+0}$ represent the coupling between the two subspaces. We can write the Schr\"odinger equation limited to the neutral meson subspace by adding to the Hamiltonian $H^0$ an effective, non-local, and energy-dependent Hamiltonian \cite{Bose-Einstein} 
\begin{equation}
 H^{0\,\prime}=H^{0+}(E-H^++i\varepsilon)^{-1}H^{+0} \,,
\end{equation}
where $E$ is the energy of the eigenstate.\footnote{The theory just described is the same one used for Feshbach resonances.} This term essentially acts as a non-local potential that adds to the delta potential of the neutral meson interactions, resulting in a non-zero effective range. This is very much along the same line as the pion contribution to $r_0$, as computed in~\cite{Esposito:2023mxw}.

We saw that under the assumption of a pure molecule (scattering from a Dirac delta alone) $r_0\geq0$ and of the order of the inverse of the mass of the mediator that has been integrated out (as in the deuteron); in the presence of an elementary $X$, $r_0<0$ and possibly large in magnitude. In Section~\ref{correpion}, it was also discussed how the pion exchange could mimic, without actually succeeding, the effect of an elementary $X$, pulling $r_0$ toward negative values even in the pure molecular hypothesis. Now we have obtained that even the presence of the charged threshold leads to a negative correction to $r_0$. Providing an estimate of this correction is complicated due to the unknown nature of the interaction parameters.

We summarize in Tab.~\ref{tab:r0} the results obtained so far for $r_0$.
\begin{table}[ht]
 \centering
 \begin{tabular}{|c|c|c|c|}
 \hline
 & Molecule & Molecule + Pion exchange & Molecule + Charged threshold \\
 \hline\hline
 $r_0$ & $\geq 0$ & $-0.20\lesssim \text{Re}\,r_0\lesssim-0.15\,\text{fm}$ & $<0$ \\
 \hline
 \end{tabular}
 \caption{Summary of the various factors that determine the sign of $r_0$. Recall that in the compact model, $r_0\lesssim-1.6\,\text{fm}$.
 }
 \label{tab:r0}
\end{table}

\subsection{Weinberg's $r_0$ directly from LHCb data}
Consider the formula for the amplitude $f$ given in~\eqref{err0formw}, with the addition of the width $\Gamma$ as in~\eqref{f11} 
\be
f(E)= -\frac{1}{m(R_0-r_0)}\frac{1}{E+B+i\tfrac{\Gamma}{2}} \,,
\label{sopra}
\ee
where $R_0=(2mB)^{-1/2}$. The residue at the pole corresponds to $(m/2\pi)g_Z^2$, see~\eqref{328} and~\eqref{330}, where $g_Z$ is the coupling of the bound state ${\mathfrak B}$ to $D\bar D^*$ in the presence of a compact state $\mathfrak X$ (which couples to $D\bar D^*$ with a coupling $g_c\neq g_Z$). 

As commented above, this formula is valid only at very low energies, in the vicinity of the shallow bound state. In the narrow region of $E$ close to the pole there would be no need to resort to Flatt\'e or to purely molecular parametrizations. The Weinberg's $r_0$ could be extracted directly fitting the LHCb data points in the energy region where $f(E)$ in~\eqref{sopra} is valid. 

Performing a fit using the LHCb data set from 2012 with $p_{\pi^+\pi^-} < 12\,\text{GeV}$ \cite{LHCb:2020xds}, i.e.  selecting only the resulting seven points around the peak, we noticed that the value of $r_0$ strongly depends on the few basic assumptions we need to treat data. 

Since $B$ is small, determining the precise position of the peak itself is challenging, given the poor statistics at hand: we expect the $X$ to lie below threshold, since Eq.~\eqref{sopra} is valid only if $B>0$. 

In conclusion, it is impossible to make any precise statement about Weinberg's $r_0$ as long as the dataset  at very low energy is so small. It is this circumstance that forces to use data at higher energies and consequently model  the tails of $f$ in the Flatt\'e or purely molecular ways; this in turn implies the definite constraints on $r_0$ we discussed above.

On top of that, we also highlight the fact that, even if a precise determination of $r_0$ were available, the charge threshold can contribute to pushing it towards negative values, as shown in Eq.~\eqref{eq:as_r0}. If one wants to rigorously disentangle the contribution from the charge threshold from that of the elementary $X$, one must follow the procedure described in Eq.~\eqref{eq:r0prime}. This in turns requires the knowledge of the couplings $\lambda_S$ and $\lambda_T$, thus preventing from drawing completely model-independent conclusions. Nonetheless, by dimensional analysis we know that the Lagrangian couplings must be of the order $\lambda_S \sim \lambda_T \sim 1/\Lambda^2$, with $\Lambda \sim 1$~GeV, and up to adimensional coefficients. One can easily check that for generic values of these coefficients, the resulting (negative) contribution $r_0^M$ is of order $1$~fm or less. The only way to make it sizable is to tune the parameters in order for the denominator in Eq.~\eqref{eq:as_r0} to vanish, thus requiring one further tuning in this already very tuned system.

In light of everything discussed above, one should also revisit the interpretation of other determinations of $r_0$, like the one extracted from lattice QCD (see, for example,~\cite{Sadl:2024dbd,Meng:2024kkp,Collins:2024sfi,Bicudo:2022cqi}).

\section{Radiative decays of $X$ and its molecular interpretation}\label{raddec}
The universal wavefunction~\eqref{universal}
\be
\Psi_{\rm mol. }(r)=\left(\frac{2mB}{4\pi^2}\right)^{1/4}\frac{\exp\big(-r\sqrt{2mB}\big)}{r} \,,
\label{universal2}
\ee 
which is the basis of the molecular interpretation of the $X(3872)$, sharply conflicts with another remarkable experimental determination, again by the LHCb collaboration: the observed ratio of branching fractions \cite{LHCb_radiative} 
\be
 \mathcal{R}_{\text{exp}}=\text{Br}(X\to\psi^\prime\gamma)/\text{Br}(X\to\psi\gamma)=1.67\pm0.21\pm0.12\pm0.04 \,.
 \label{radi}
\ee
If the decay dynamics of the $X$ into $\psi^\prime$ and $\psi$ is the same, we would have rather expected ${\cal R}<1$, due to an obvious phase space argument. We will try to understand to which extent the molecular picture can justify this result. 

\begin{figure}[t]
 \centering
 \includegraphics[width=0.36\textwidth]{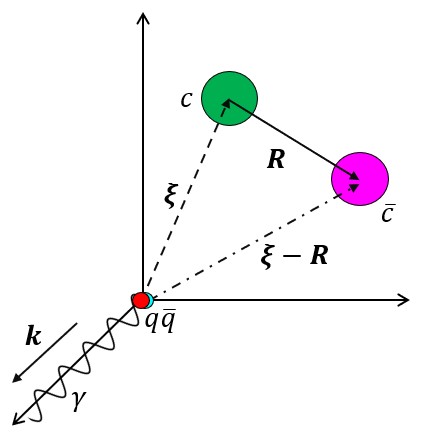}
 \caption{Scheme of the dominant process for the radiative decay $X\to\psi^{(\prime)}\gamma$. Light quarks annihilate at the origin producing a photon with momentum $\bm k$.}
 \label{fig:annhil}
\end{figure}

At the lowest order, the process $X\to\psi^{(\prime)}\gamma$ is dominated by the annihilation of the $q\Bar{q}$ pair. Without loss of generality, we assume that the annihilation takes place in the origin of the frame in Fig.~\ref{fig:annhil}. Defining $\psi(|\boldsymbol{R}|)$ as the wave function of the final charmonium, as a function of the distance between the $c\bar c$ pair, $\bm R$, the transition amplitude $A$ in the rest frame of the $X$, at a fixed photon three-momentum $\bm k$, is given by~\cite{Grinstein:2024rcu}\footnote{It has been shown \cite{Grinstein:2024rcu} that only the real part of the exponential factor contributes to the amplitude
\be
 A\left(X\to\psi^{(\prime)}\gamma\right)=\mathcal{F}\int_{\mathbf{R},\boldsymbol{\xi}} \cos\left[k\left(\cos\lambda\left(\frac{R}{2}-r\cos\theta\right)-r\sin\theta\sin\lambda\cos\phi\right)\right]
 \psi_c(|\boldsymbol{R}|)\Psi_{c\Bar{c}}(|\boldsymbol{R}|)\Psi_{q\Bar{q}}(|\boldsymbol{\xi}|,|\boldsymbol{\xi}-\boldsymbol{R}|)\notag \,.
\ee
The final $c\bar c$ pair recoils against the photon with non-relativistic velocity $\bm V=\bm k/2M$, where $k\bm $ is given in \eqref{kappar}
\begin{align}
\psi\to (\Phi_{\boldsymbol x}\,, \, e^{-i\boldsymbol K\cdot \bm V}\Psi)&=\int_{\boldsymbol p}(\Phi_{\boldsymbol x}\, ,\, \chi_{\boldsymbol p})(\chi_{\boldsymbol p}\, ,\, e^{-i\boldsymbol K\cdot \bm V}\Psi)=\int_{\boldsymbol p}(\Phi_{\boldsymbol x}\, ,\, \chi_{\boldsymbol p})(\chi_{\boldsymbol p-2M\boldsymbol V}\, ,\, \Psi)=\notag\\
& = \int_{\boldsymbol p}(\Phi_{\boldsymbol x}\, ,\, \chi_{\boldsymbol p + 2M\boldsymbol V})(\chi_{\boldsymbol p},\Psi)=e^{i\boldsymbol k\cdot \boldsymbol x}\, \psi \notag \,,
\end{align}
with $\boldsymbol x=\big(\boldsymbol \xi +\boldsymbol \eta\big)/2$ and $\bm \eta=\bm \xi-\bm R$.}
\be
 A\left(X\to\psi^{(\prime)}\gamma\right)=\mathcal{F}\int_{\bm R,\bm\xi} e^{-i\bm k\cdot\left(\boldsymbol{\xi}-\frac{\boldsymbol{R}}{2}\right)}\,\psi(|\boldsymbol{R}|)\Psi_{c\Bar{c}}(|\boldsymbol{R}|)\Psi_{q\Bar{q}}(|\boldsymbol{\xi}|,|\boldsymbol{\xi}-\boldsymbol{R}|) \,,
 \label{eq:A}
\ee
where the term 
\be
\Psi_{c\Bar{c}}(|\boldsymbol{R}|)\Psi_{q\Bar{q}}(|\boldsymbol{\xi}|,|\boldsymbol{\xi}-\boldsymbol{R}|) \,,
\ee
is factorized as in 
\be
\Psi_{\rm mol.}(|\boldsymbol{r}_c-\boldsymbol{r}_{\bar c}|)\, \Xi_M(|\boldsymbol{r}_u-\boldsymbol{r}_{\bar c}|)\,\Xi_M(|\boldsymbol{r}_{\bar u}-\boldsymbol{r}_{c}|)\equiv \Psi_{\rm mol.}(R)\Xi_{D}\Xi_{\bar D^*} \,,
\label{lamolecola}
\ee
where $\Psi_{\rm mol.}(R)$ is the solution of the $\delta^3(\bm R)$ potential binding two pointlike $D$ and $\bar D^*$ mesons into a $X(3872)$ molecule whereas the $\Xi_M$'s represent the wavefunctions of the light quarks in $D$ and $\bar D^*$ extended mesons \cite{Isgur:1988gb}. The factor $\mathcal{F}$ takes into account various normalizations. 

The value of $\mathcal{R}$ depends only on the ratio of the squared moduli of the amplitudes \eqref{eq:A}, the ratio of phase spaces $\Phi=0.26$, and the sum over polarizations $\mathcal{P}=0.98$, which, depending on the momentum $\bm k$ of the produced photon 
\be
 |\bm k|=\frac{M_X^2-M^2_{\psi^{(\prime)}}}{2M_X} \,,
 \label{kappar}
\ee
do not cancel out in the ratio. Putting everything together yields the compact formula
\be
\mathcal{R}=\Phi\,\mathcal{P}\left|\frac{A\left(X\to\psi^{\prime}\gamma\right)}{A\left(X\to\psi\gamma\right)}\right|^2\simeq \frac{1}{4}\left|\frac{A\left(X\to\psi^{\prime}\gamma\right)}{A\left(X\to\psi\gamma\right)}\right|^2 \,.
\label{eq:Rteorica}
\ee

In place of $\Xi_M$ above we use for $D$ and $\bar D^*$
\be
\Xi_M(r)=\frac{b^{3/2}}{\pi^{3/4}}e^{-\frac{1}{2}b^2 r^2} \,,
\ee
with the parameter $b$ as in~\cite{Isgur:1988gb}, defining a size of open charm mesons of about
\be
\sqrt{\langle r^2\rangle_{\Xi_M}}\simeq 0.68~\text{fm} \,.
\ee
This way we obtain 
\be
{\cal R}\simeq 0.034 \,,
\label{rmolx}
\ee
which cannot be reconciled with the value in~\eqref{radi}. 

To overturn the situation one might assume that:
\begin{enumerate}
\item $D$ mesons are as large as 0.9~\text{fm} or more, which translates into a different value of $b$, in contrast with what found in~\cite{Isgur:1988gb}. This goes in the direction of enlarging the probability of two light quarks to meet and annihilate into a photon --- remind that the loosely bound molecule is a large object with $R_0\simeq 14~\text{fm}$ for $B=100$~keV, with $D$ mesons, and the light quarks in them, being far apart on average. 
\item The $\Psi_{\rm mol.}(R)$ is not an adequate description at very short distances.
\end{enumerate}
Recall that $\Psi_{\rm mol.}(R)$ does not depend on the details of the potential featuring the shallow bound state (and it is therefore named as {\it universal} wavefunction). Since $\Psi_{\rm mol.}(R)$ is very broad, for a shallow bound state, we can approximate the undefined potential $V$ as an attractive Dirac delta, which, upon renormalization of the coupling, can have a bound state. Indeed the bound state wavefunction in the $\delta^3(\bm R)$ potential can be shown to have exactly the form in~\eqref{universal2}. 

It is certainly true that one-pion exchange interaction could modify the short distance behavior and lead to a different $\Psi_{\rm mol.}$ at short distance. As discussed in Section~\ref{corrs}, the potential due to the one-pion exchange is (see Eq.~\eqref{weak})
\be
V_w(r)=-\alpha\frac{e^{i\mu r}}{r} \,,
\ee
where
\be
\alpha=\frac{g^2\mu^2}{24\pi f_\pi^2}=5\times 10^{-4} \,, \qquad\qquad f_\pi\simeq 132~\text{MeV} \,,
\ee
and 
\be
\mu=\sqrt{2m_\pi \delta}\simeq 43~\text{MeV} \,, \qquad\qquad \delta=m_{D^*}-m_{D}-m_\pi \,.
\ee
The correction to $\Psi_{\text{mol.}}$ is not expected to be large enough to considerably modify $\mathcal{R}$ since it will be proportional to $\alpha=5\times10^{-4}$. Indeed, it can be computed in perturbation theory at first order
\be
\Psi_{\rm mol.}^{(1)}=\Psi_{\rm mol.}+\delta_1 \Psi_{\rm mol.} \,,
\ee
where \cite{Weinberg_QFT}
\be
\delta_1\Psi_{\rm mol.}=-\int_0^{+\infty} \Psi_s \frac{\Big(\Psi_s,V_w \Psi_{\rm mol.}\Big)}{B+E}\,\rho(E)\, dE \,,
\ee
with (see Eq.~\eqref{eq:phase_space})
\be
\rho(E)=\frac{1}{4\pi^2}(2m)^{3/2}\sqrt{E} \,.
\ee
The scattering wavefunction in the Dirac potential, as computed in~\cite{jackiw}, is
\begin{equation}
\Psi_{s}(R, k)=e^{i\delta(k)}\frac{\sin(k R+\delta(k))}{kR} \,, \qquad\text{with}\qquad k=\sqrt{2mE} \,.
\label{eq:psi_s}
\end{equation}
The $S$-wave phase shift $\delta(k)$ is given by (see footnote~\ref{footx})
\be
\delta(k)=\arctan(-a_s\,k) \,,
\ee
where the scattering length $a_s$ is linked to $B$ by \eqref{eq:a_B}
\begin{equation}
\frac{1}{a_s}=\sqrt{2mB} \,.
\end{equation}
Thus, the universal molecular wavefunction $\Psi_{\text{mol.}}(R)$ \eqref{universal2} can be rewritten as
\begin{equation}
\Psi_{\text{mol.}}(R)=\frac{1}{\sqrt{2\pi a_s}}\frac{e^{-R/a_s}}{R} \,.
\end{equation}

It is slightly more convenient to use the reduced wave functions, $u_{\rm mol.}^{(1)}= u_{\rm mol.}(R)+\delta_1 u_{\rm mol.}(R)$, so that
\begin{equation}
u_{\rm mol.}^{(1)} (R) =\sqrt{\frac{2}{a_s}}e^{-\frac{R}{a_s}}-\int_{0}^{+\infty}\,\frac{\sin\big[R\sqrt{2m E}+\delta(E)\big]}{\sqrt{2m E}}\frac{A(E)}{B+E}\,\rho(E)\,dE \,,
\label{eq:chiboundpert}
\end{equation}
where
\begin{equation}
A(E)=-4\pi\sqrt{\frac{2}{a_s}}\alpha\int_{0}^{+\infty} \frac{\sin\big[r\sqrt{2mE}+\delta(E)\big]}{\sqrt{2mE}}\,\frac{e^{i\mu r}}{r}\,e^{-\frac{r}{a_s}}\,dr \,.
\end{equation}
The integral has a logarithmic ultraviolet divergence for $r\to0$. To regularize it, we introduce a short-distance cutoff, $\eta$
\begin{equation}
A(E)=-4\pi\sqrt{\frac{2}{a_s}}\alpha\int_{\eta}^{+\infty} \frac{\sin\big[r\sqrt{2mE}+\delta(E)\big]}{\sqrt{2mE}}\,\frac{e^{i\mu r}}{r}\,e^{-\frac{r}{a_s}}\,dr\,.
\label{25}
\end{equation}
Expanding~\eqref{25} for small $\eta$, a logarithmic divergence appears in $A(E)$
\begin{equation}
A(E)=\frac{4\pi}{\sqrt{2mE}}\alpha\sqrt{\frac{2}{a_s}}\sin\delta(E)\,\log(\eta\mu)+\text{finite} \,,
\label{eq:EBlog}
\end{equation}
so that
\begin{equation}
u_{\rm mol.}^{(1)} (R)=\sqrt{\frac{2}{a_s}}e^{-\frac{R}{a_s}}-\alpha\, m\sqrt{\frac{2}{a_s}}e^{-\frac{R}{a_s}}(a_s-2R)\log(\eta\mu)+\text{finite} \,.
\label{eq:u_mol_div}
\end{equation}
This might suggest that the perturbative correction to $u_{\rm mol.}(R)$ contains an ultraviolet divergence that could invalidate the result. The method to absorb this divergence is provided in \cite{Esposito:2023mxw}. In this work, it is presented an alternative approach (with respect to the one discussed in Section \ref{corrs}) to regularize the divergences that arise when attempting to compute the corrections to the scattering amplitude $f$ due to $V_w$. The key point is to treat $a_s$ as the bare scattering length and introduce a renormalized scattering length $a$.\footnote{This is equivalent to stating that we should write \eqref{universal2} using a bare binding energy $B_0$, while $B$ is the finite and experimentally measurable binding energy.} In the absence of pion interactions
\begin{equation}
a_s=a=\frac{1}{\sqrt{2mB}} \,.
\end{equation}
If we require that the $O(k^0)$ term of $f^{-1}$ in \eqref{f_inv} computed with \eqref{eq:psi_s} is finite and equal to $a$, we obtain by expanding at $O(\alpha)$ \cite{Esposito:2023mxw}
\begin{equation}
a_s=a\left(1-2\frac{\alpha m}{\mu}\left(\frac{1}{a\mu}+\gamma_E\, \mu\, a+2i+\mu a\left(\log(\eta\mu)-i\frac{\pi}{2}\right)\right)\right) \,,
\label{eq:asaR}
\end{equation}
where $\gamma_E$ is the Euler-Mascheroni constant.

Consistently, we should write \eqref{eq:u_mol_div} as
\begin{equation}
u_{\rm mol.}^{(1)} (R)=\sqrt{\frac{2}{a_s}}e^{-\frac{R}{a_s}}-\alpha\, m\sqrt{\frac{2}{a}}e^{-\frac{R}{a}}(a-2R)\log(\eta\mu)+\text{finite} \,,
\end{equation}
since the second term is already $O(\alpha)$. Using \eqref{eq:asaR} to expand the first term gives
\begin{equation}
\sqrt{\frac{2}{a_s}}e^{-\frac{R}{a_s}}=\sqrt{\frac{2}{a}}e^{-\frac{R}{a}}+\alpha\, m\sqrt{\frac{2}{a}}e^{-\frac{R}{a}}(a-2R)\log(\eta\mu)+\text{finite} \,.
\label{eq:chidiverg}
\end{equation}

This shows that the divergent term gets canceled at order $\alpha$. Including all the finite terms, we get
\be
u_{\rm mol.}^{(1)} (R)=\sqrt{\frac{2}{a}}e^{-\frac{R}{a}}+\alpha\,\Re (F(R))+i\alpha\,\Im (F(R)) \,,
\ee
where
\begin{equation}
F(R)=\frac{m}{2a^2\,\mu^2}(a-2R)\left[2+a\mu\left(4i+a\mu(2\gamma_E-i\pi)\right)\right]\,\sqrt{\frac{2}{a}}\,e^{-\frac{R}{a}}+\bar A(E) \,,
\label{eq:intfinitecorr}
\end{equation}
and $\bar{A}(E)$ is the finite part of the integral \eqref{25} after factoring out $\alpha$.

The value ${\cal R}$ can be recomputed with the aid of $\Psi^{(1)}_{\rm mol.}$, giving
\begin{equation}
\mathcal{R}=0.043 \,,
\label{terzaz}
\end{equation}
not qualitatively different from~\eqref{rmolx}.
The qualitative agreement between these two results seem like an unmistakable sign of an elementary $X$.

\section{Conclusions}
The radiative decays of the $X(3872)$ are not compatible with the basic version of its shallow
bound state interpretation. In addition the available LHCb data analysis of the $X(3872)$ lineshape only tells that observations are compatible, to some extent, with a compact tetraquark state, the excitation of a $X$ field coupled to the $D$ and $\bar D^*$ meson fields in an effective Lagrangian description. We propose a second fit to a purely molecular lineshape in order to appreciate the difference between the two on the basis of present data. 

There are various recent attempts to describe the $X$ with the methods of the Born-Oppenheimer approximation, where the charm quarks substitute the heavy nuclei in ordinary molecules, held together by the Born-Oppenheimer potential generated by the quark motion~\cite{BOold1,BOold2,BOold3,BOrattz}. 
In~\cite{Germani:2025qhg} the $X,Z_c(3900)$ and $X(4020)$ particles can be described as compact tetraquarks whose quantum state at large distance is a superposition $(c\bar q)_{\bm 1}(\bar c q)_{\bm 1}+(c\bar q)_{\bm 8}(\bar c q)_{\bm 8}$, rather than $\bm{3},\bar{\bm{3}}+\bm{6},\bm{\bar{6}}$, which instead is found to produce much heavier particles. Studies of this kind are being proposed also by other groups~\cite{brambi0,brambi,bruschini}. The outcome is very similar to the one reached above: the purely molecular description of the $X$, and of similar hadrons, is inadequate and even if the original proposal for a compact $X$~\cite{Maiani:2004vq} is challenged by data, it has still to be taken into account when facing the construction of a quantitative interpretation of exotic hadrons. 

Although the definition of a particle as a physical system identified by only {\it one continuous} degree of freedom, its free momentum $\bm p$, has no ambiguities, deciding if that particle is elementary or not is more problematic. In the effective field theory context, more levels of elementariness are possible: quarks and gluons do not appear in the effective theory description of nucleon interactions at low energies, but are associated to the fields of quantum chromodynamics. In this sense quarks and gluons are more elementary than nucleons and mesons, whereas nuclei are less elementary than nucleons. In this review we addressed the problem of determining the degree of elementariness of the $X(3872)$ exotic hadron, and possibly of its very many similar states: is the $X$ as elementary as $D$,$\bar D^*$ mesons, or less? Our conclusion, based on the present literature, is that the $X$ is not less elementary than open-charm mesons. More tests of this conclusion are needed, and we hope that part of them are described in this work. 

\section*{Acknowledgements}
We thank V. Belyaev, B. Grinstein, D.~B. Kaplan,  L. Maiani, A. Pilloni, R. Rattazzi for various discussions and collaboration on the topics described in this review. The work of A. G. was supported in part by the European Union - Next Generation
EU under italian MUR grant PRIN-2022-RXEZCJ.

\bibliographystyle{hacm}
\bibliography{biblio.bib}
% \printbibliography

\end{document}